\documentclass{aa}  

\usepackage{graphicx}
\usepackage{color}
\usepackage{multirow}
\usepackage{natbib}
\usepackage{savefnmark}
\bibpunct{(}{)}{;}{a}{}{,}
\usepackage{txfonts}
\begin{document}

   \title{The \object{NGC~1614} interacting galaxy}

   \subtitle{Molecular gas feeding a ``ring of fire''}

   \author{S. K\"onig
          \inst{1,2}
          \and
          S. Aalto
	   \inst{3}
	  \and
          S. Muller
	   \inst{3}
	  \and
          R.~J. Beswick
	   \inst{4}
	  \and
          J.~S. Gallagher III
	   \inst{5}
          }

   \institute{Institut de Radioastronomie Millim\'etrique, 300 rue de la Piscine, Domaine Universitaire, 38406 Saint 
  	     Martin d'H\`eres, France\\
              \email{koenig@iram.fr}
	 \and 
	     Dark Cosmology Centre, Niels Bohr Institute, University of Copenhagen,
              Juliane Maries Vej 30, 2100 Copenhagen, Denmark\\
         \and
             Chalmers University of Technology, Department of Radio and Space Science, Onsala Space Observatory, 43992 
	     Onsala, Sweden\\
         \and
             University of Manchester, Jodrell Bank Centre for Astrophysics, Oxford Road, Manchester, M13 9PL, UK\\
         \and
             Department of Astronomy, University of Wisconsin, 475 N. Charter Street, Madison, WI, 53706, USA\\
             }

   \date{Received ; accepted }

 
\abstract
{Minor mergers frequently occur between giant and gas-rich low-mass galaxies and can
 provide significant amounts of interstellar matter to refuel star formation and power
 active galactic nuclei (AGN) in the giant systems. Major starbursts and/or AGN result when
 fresh gas is transported and compressed in the central regions of the giant galaxy. %
 This is the situation in the starburst minor merger \object{NGC~1614}, whose molecular
 medium we explore at half-arcsecond angular resolution through our observations of 
 $^{\rm 12}$CO\,(2$-$1) emission using the Submillimeter Array (SMA). We compare our 
 $^{\rm 12}$CO\,(2$-$1) maps with optical and Pa$\alpha$, \textit{Hubble Space Telescope}
 and high angular resolution radio continuum images to study the relationships between dense
 molecular gas and the NGC~1614 starburst region. The most intense $^{\rm 12}$CO emission 
 occurs in a partial ring with $\sim$230~pc radius around the center of NGC~1614, with an
 extension to the northwest into the dust lane that contains diffuse molecular gas. %
 We resolve ten giant molecular associations (GMAs) in the ring, which has an integrated
 molecular mass of $\sim$8\,$\times$\,10$^{\rm 8}$~M$_{\sun}$. Our interferometric
 observations filter out a large part of the $^{\rm 12}$CO\,(1$-$0) emission mapped at
 shorter spacings, indicating that most of the molecular gas is diffuse and that GMAs only
 exist near and within the circumnuclear ring. The molecular ring is uneven with most of
 the mass on the western side, which also contains GMAs extending into a pronounced tidal
 dust lane. %
 The spatial and kinematic patterns in our data suggest that the northwest extension of the
 ring is a cosmic umbilical cord that is feeding molecular gas associated with the dust lane 
 and tidal debris into the nuclear ring, which contains the bulk of the starburst activity. The
 astrophysical process for producing a ring structure for the final resting place of
 accreted gas in NGC~1614 is not fully understood, but the presence of numerous GMAs
 suggests an orbit-crowding or resonance phenomenon. There is some evidence that star
 formation is progressing radially outward within the ring, indicating that a
 self-triggering mechanism may also affect star formation processes. The net result of this
 merger therefore very likely increases the central concentration of stellar mass in the
 NGC~1614 remnant giant system.}


   \maketitle
%

\section{Introduction}

\begin{figure*}[t]
  \begin{minipage}[hbt]{0.33\textwidth}
  \hspace{-0.7cm}
  \centering
    \includegraphics[width=1.05\textwidth]{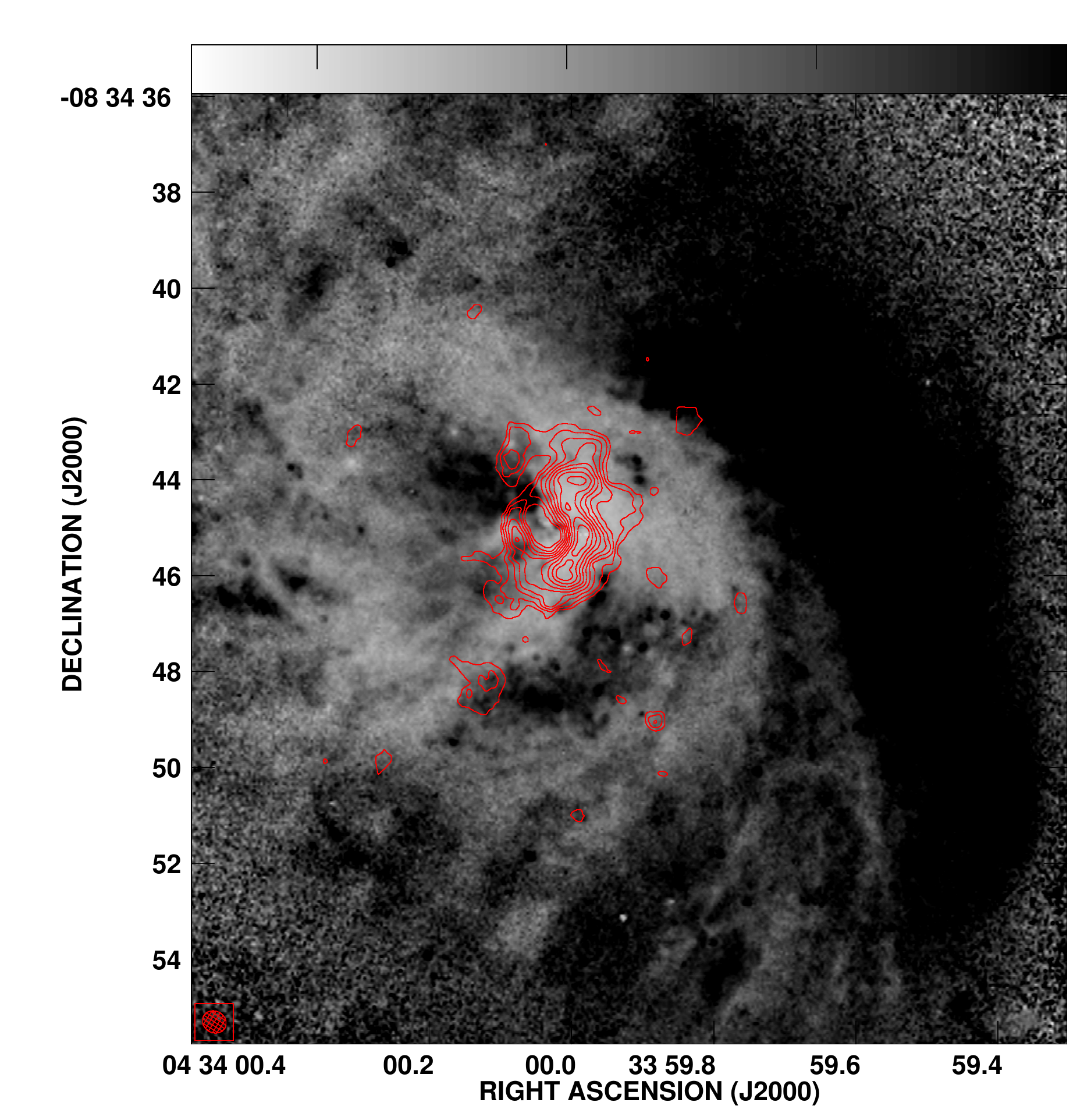}
  \end{minipage}
  \begin{minipage}[hbt]{0.33\textwidth}
  \centering
    \includegraphics[width=0.87\textwidth,angle=-90]{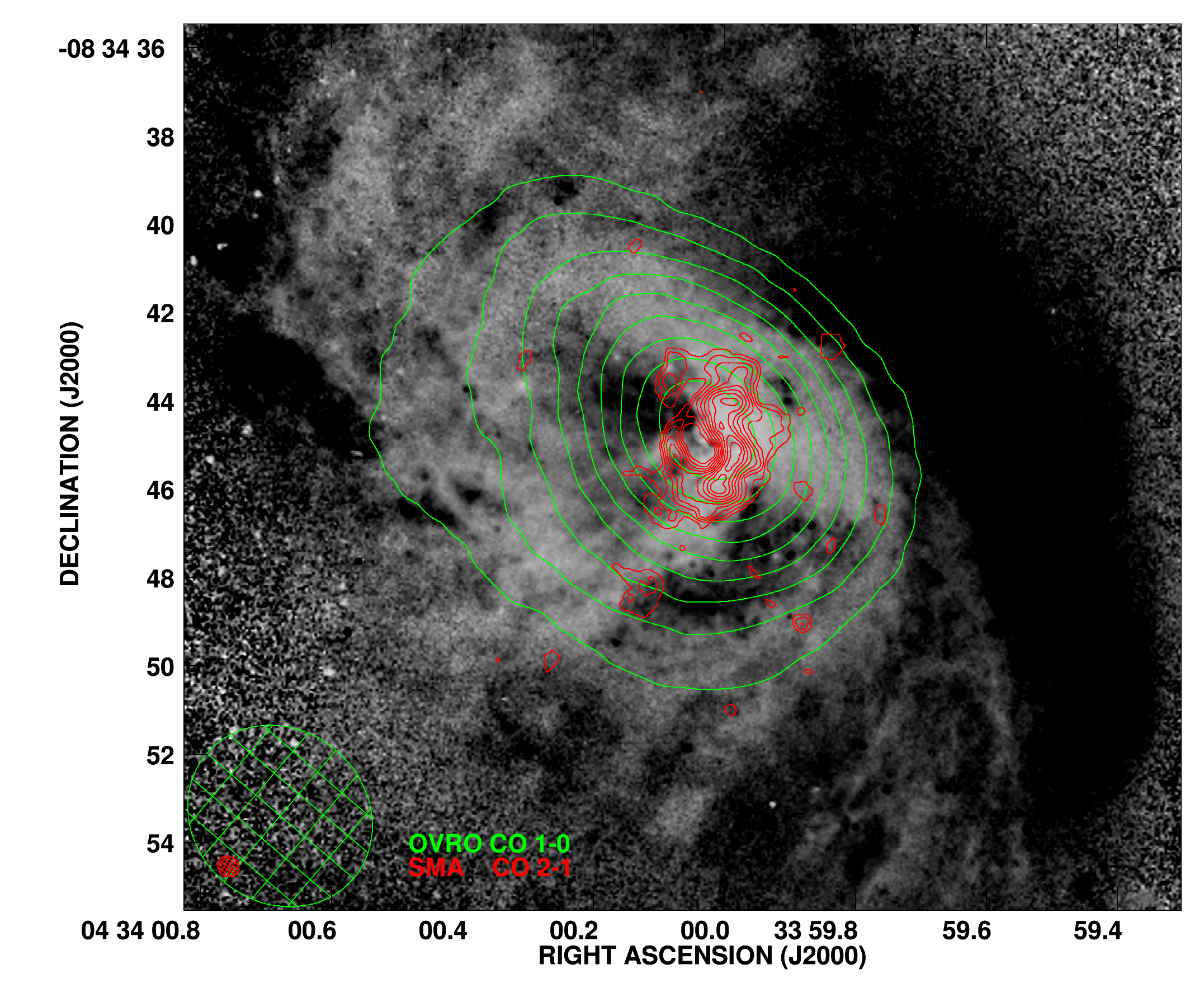}
  \end{minipage}
  \begin{minipage}[hbt]{0.33\textwidth}
  \centering
    \includegraphics[width=1.035\textwidth]{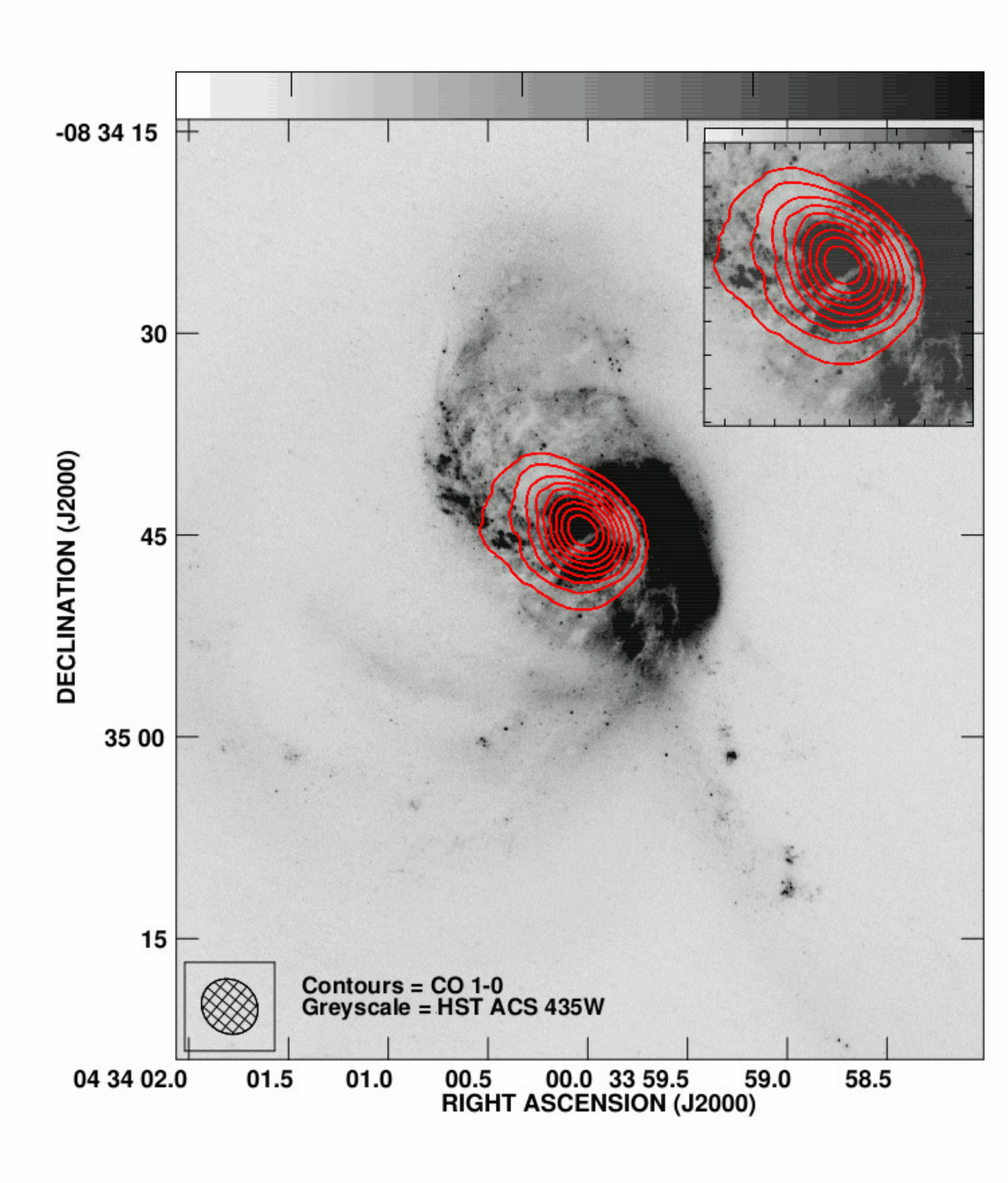}
  \end{minipage}
  \caption{\footnotesize Overlay of the CO\,(2$-$1) integrated intensity emission \citep[\textit{left},][]{ols10} and of the 
  CO\,(2$-$1) and low-resolution CO\,(1$-$0) \citep[\textit{middle},][]{ols10} integrated intensity emission on the HST F435W/F814W 
  filter color-map image, and overlay of the CO\,(1$-$0) emission (\textit{right}) on an HST F435W filter image on the scale of the 
  whole merger system. Note how well the CO\,(2$-$1) emission distribution correlates with the dust lanes (light gray colors 
  on the left-hand side and the middle in this figure) and that the position of the emission peak of the CO\,(1$-$0) distribution 
  coincides with the main part of the dust emission in this region. The CO beam sizes (CO\,(2$-$1): 0.50\arcsec\,$\times$\,0.44\arcsec, 
  CO\,(1$-$0): 4.44\arcsec\,$\times$\,4.10\arcsec) are shown in the lower left corners.}
  \label{fig:overlay_co2-1_co1-0_HST}
\end{figure*}

Merger events, i.e., close interactions of galaxy pairs, have a deep impact on the evolution of galaxies through the triggering 
of active galactic nuclei (AGN) and starburst activity. Studies of interacting galaxies often focus on major mergers (nearly equal 
mass spirals) \citep[e.g.][]{too72,tac02,das06,jes07,som08,kor09} and their evolution, although minor mergers most likely constitute 
the bulk of all galaxy interactions \citep[e.g.][]{toth92,zar95,hun96}. Investigating how the gas is feeding starburst and AGN 
activities in these objects is therefore paramount for understanding the overall evolution of the Universe. Although there is still 
no consensus what powers the emission in low ionization nuclear emission line regions (LINERs), nuclear starburst, AGN activity, 
and shocks are three major candidates \citep[e.g.][]{vei99,ho99,tera00,alo00,mon06}. Bar-driven inflow of gas in galaxies has 
been suggested to trigger nuclear starbursts, and the feeding of AGNs \citep[e.g.][]{sim80,sco85}. However, the radial inflow of gas 
along the bar may be slowed down at certain radii, which are often associated with inner Lindblad resonances 
\citep[ILR, ][]{com88a,shlo89}.\\
\indent
\object{NGC~1614} is an excellent example of a minor merger with a spectacular morphology and intense nuclear activity. The 
nucleus of this galaxy displays characteristics of containing an intense nuclear starburst region and may also contain some level 
of AGN activity. NGC~1614 \citep[SB(s)c~pec,][]{deV91} is a luminous infrared galaxy \citep[LIRG,][]{san03} at a distance 
of 64~Mpc \citep[][for $H$$_{\rm 0}$\,=\,75~km\,s$^{\rm -1}$ $\rm {Mpc^{-1}}$; 1\arcsec\,=\,310~pc]{deV91}. Most of the starburst 
activity can be attributed to the interaction with a companion galaxy located in the tidal tail to the southwest of the nucleus of 
NGC~1614 \citep{neff90,ols10,vai12}. A mass ratio of $\sim$5:1\,$-$\,3:1 has been found for this minor merger system \citep{vai12}. 
Its nuclear optical spectrum shows both starburst and LINER activity. A circumnuclear ring ($r$\,=\,300~pc, see e.g. 
Fig.\,\ref{fig:overlay_co2-1_co1-0_HST}) in NGC~1614, seen in tracers such as Pa$\alpha$ \citep[][along with an identification in 
$H-K$ imaging]{alo01}, radio continuum \citep{ols10} and polycyclic aromatic hydrocarbons (PAHs) \citep{vai12}, hosts a very young 
starburst \citep[5--10~Myr, top-heavy initial mass function,][]{pux99} while an older ($>$\,10~Myr) starburst resides in the 
nucleus - possibly encircling an AGN. 
\citet{alo01} claimed that the ring is the cause of ``wildfire'' starburst activity propagating from the nucleus in an outward 
direction through cloud-cloud collisions and winds, while \citet{ols10} instead suggest that the ring is the result of a resonance, 
possibly linked to a stellar bar.\\
\indent
Though NGC~1614 has intense nuclear activity, its spiral arms are devoid of molecular gas, which raises the question how the gas 
reached the nucleus. Numerical simulation studies of minor- or intermediate mergers show that gas brought in by the disturbing 
companion galaxy is generally found at large radii in the merger remnant \citep{bour05}. Gas returns to the system from tidal tails 
and often forms rings - polar, inclined or equatorial - that will appear as dust lanes when seen edge-on.\\
\begin{figure*}[!ht]
  \begin{minipage}[hbt]{0.4925\textwidth}
  \centering
    \includegraphics[width=0.75\textwidth,angle=-90]{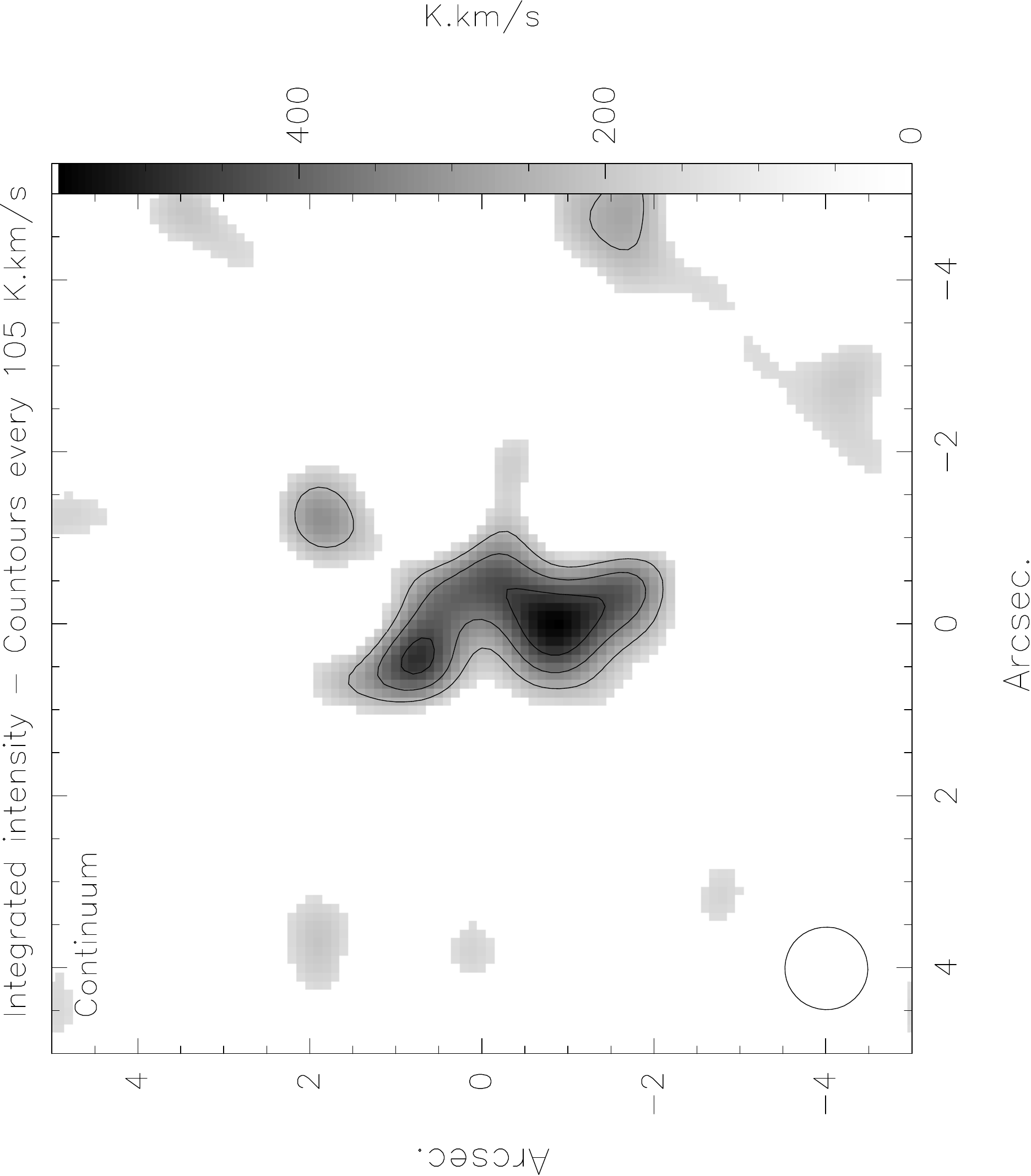}
    \includegraphics[width=0.75\textwidth,angle=-90]{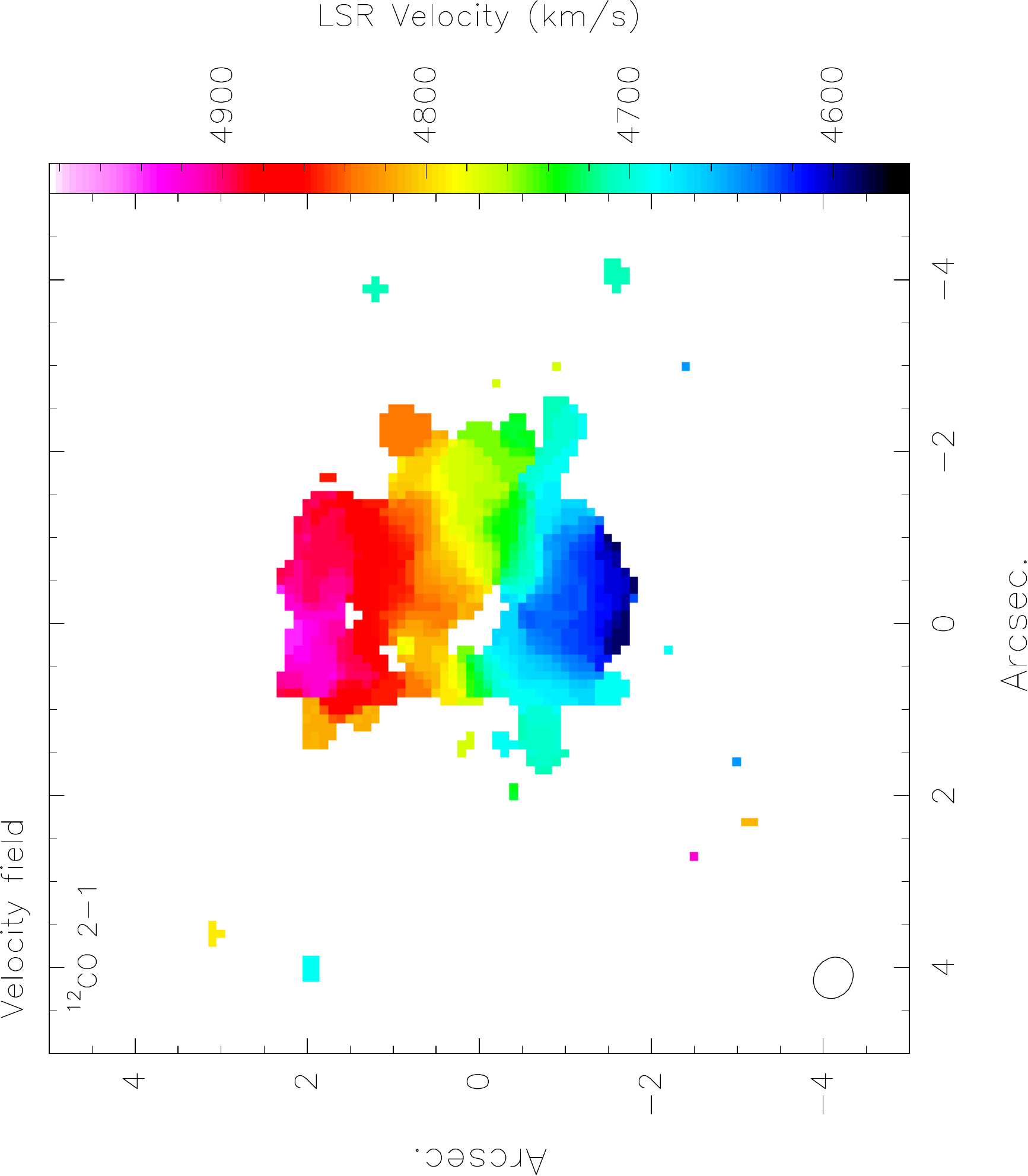}
  \end{minipage}
  \begin{minipage}[hbt]{0.4925\textwidth}
  \centering
    \includegraphics[width=0.75\textwidth,angle=-90]{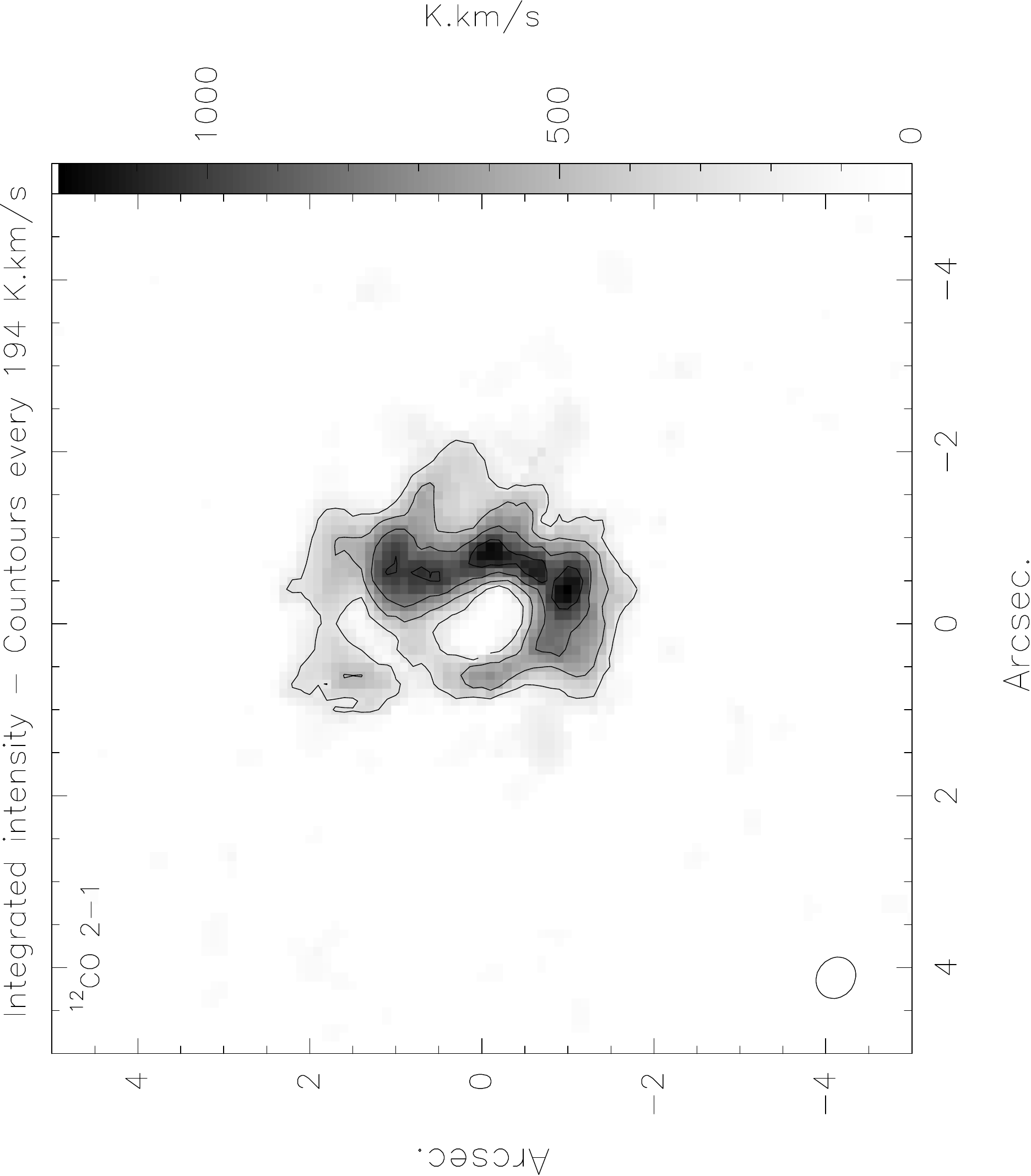}
    \includegraphics[width=0.75\textwidth,angle=-90]{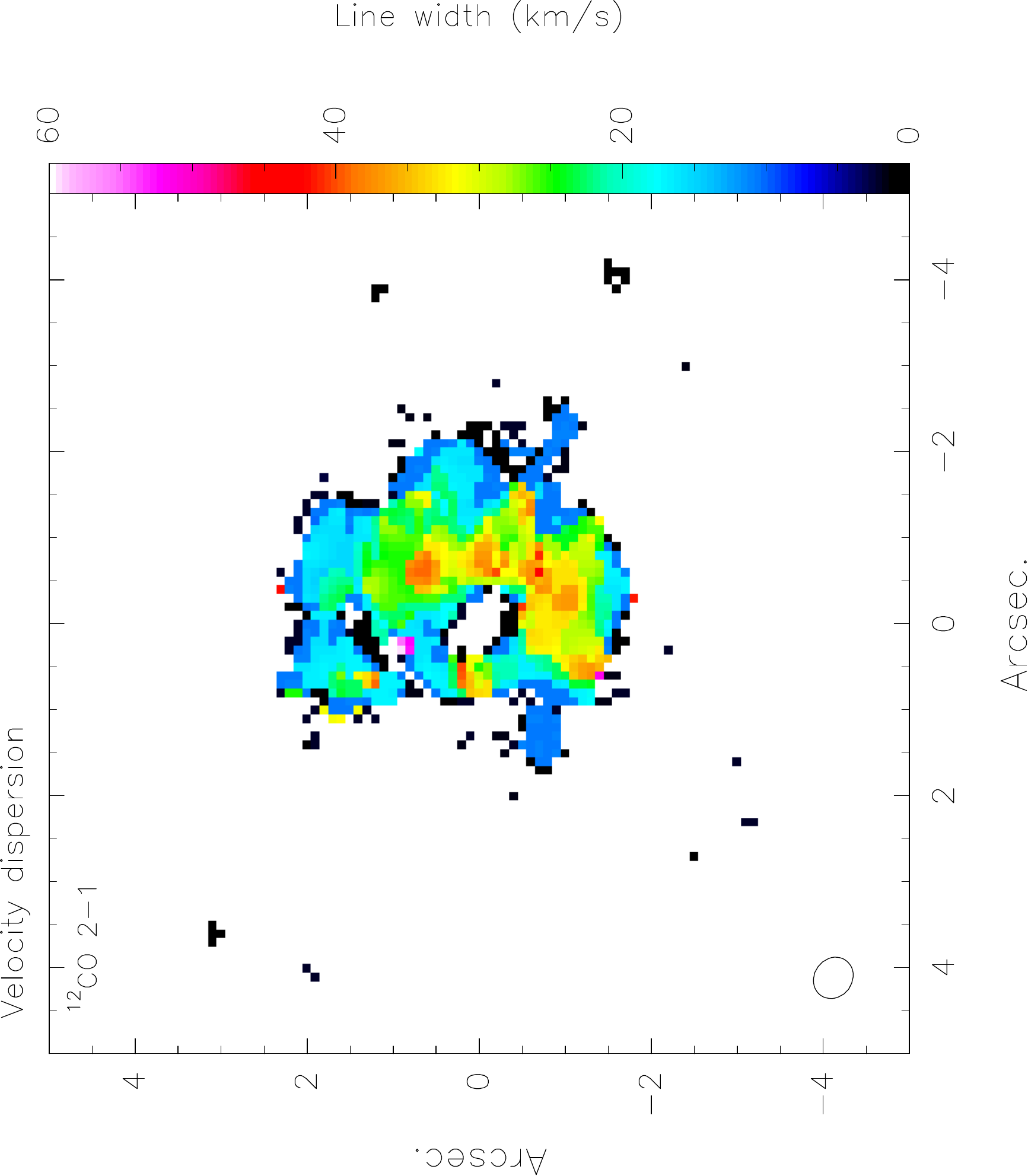}
  \end{minipage}
  \caption{\footnotesize Smoothed naturally weighted SMA 1.3~mm continuum (\textit{upper left}) in \object{NGC~1614}. The noise level 
  is 0.71~mJy/beam. Contours start at 1.5$\sigma$ in steps of 1.1$\sigma$. Integrated intensity map (\textit{upper right}), velocity 
  field (\textit{bottom left}), and velocity dispersion (\textit{bottom right}) of the CO\,(2$-$1) emission in NGC~1614. The noise 
  level is 6.16~mJy/beam. Contours for the integrated intensity map start at 4$\sigma$ in steps of 4$\sigma$, contours for the 
  velocity field and velocity dispersion start at 5$\sigma$. The velocity range plotted in the velocity field is  between 
  $\sim$4520~km\,s$^{\rm -1}$ and $\sim$4985~km\,s$^{\rm -1}$. The beam sizes are indicated in the lower left corner of each plot 
  and are listed in Sect.\,\ref{sec:obs}.}
  \label{fig:sma_cont+12co2-1}
\end{figure*}
\indent
One dominant feature of the optical morphology of NGC~1614 is the prominent dust lane that crosses the galaxy just north of 
the central activity. A considerable fraction of the CO\,(1$-$0) emission resides in this dust lane and is not directly involved 
in the nuclear activity. The OVRO CO\,(1$-$0) observations (2.5\arcsec\ resolution) presented by \citet{ols10} were unable to spatially 
resolve the molecular structure of the inner region.\\
\indent
Furthermore, a spatial correlation between CO emission and 1.4~GHz radio continuum is commonly observed in galaxies. This is not the 
case for NGC~1614, however. There is probably no star formation associated with the dust lane - possibly because gas is being funneled 
along it, with the resulting shear, which prevents efficient star formation. A similar CO-dust lane correlation has been found for 
another minor merger, the \object{Medusa merger}, for which the gas in the dust lane is suggested to be diffuse because of the 
gas-funneling \citep{aalto00}.\\
\indent
Here we present a study of the $^{\rm 12}$CO emission toward the center of NGC~1614. In Sect.\,\ref{sec:obs} 
we describe the observations and data reduction, and in Sect.\,\ref{sec:results} we present the results. 


\section{Observations} \label{sec:obs}

\object{NGC~1614} was observed with the Submillimeter Array (SMA) in the $^{\rm 12}$CO\,(2$-$1) and $^{\rm 13}$CO\,(2$-$1) 
transitions. We obtained one full track (10~hr), between 22 LST and 08 LST, in the SMA's most extended 
configuration, the Vext array configuration, with a resolution of 0.5\arcsec on 26 July 2010, which provided baselines between 
68~m up to $\sim$509~m. Our map is therefore sensitive to structures smaller than 2.4\arcsec. Using the 230~GHz band 
receiver in conjunction with the DR\,128 correlator (DR\,=\,default resolution) mode with a bandwidth of 4~GHz provided channel 
widths of 0.8125~MHz, resulting in a velocity coverage of 5305.7~km\,s$^{\rm -1}$ with a channel separation of 
1.1~km\,s$^{\rm -1}$. For data analysis a final channel width of 21~km\,s$^{\rm -1}$ was used. The 4~GHz band 
was centered at a rest frequency of 226.931~GHz (corresponding to a central velocity of 
$v$$_{\rm opt,hel}$\,=\,4778~km\,s$^{\rm -1}$) for the upper side band (USB). The system noise temperature $T$$_{\rm sys}$ ranged between  
80 and 170~K. During the observations several sources were used as calibrators: \object{3C454.3} 
as the band pass calibrator, the nearby source \object{0423-013} for the phase calibration of NGC~1614, and 
\object{Neptune} and \object{Callisto} as flux calibrators.\\
\indent
We reduced and analyzed the data using the MIR IDL and GILDAS\footnote{http://www.iram.fr/IRAMFR/GILDAS} Mapping data 
reduction software packages. For the CO\,(2$-$1) line the synthesized beam resulting from natural weighting is 
0.50\arcsec\,$\times$\,0.44\arcsec\ with a position angle of 54\degr. For the 1~mm continuum observations the smoothed 
synthesized beam from natural weighting is 0.96\arcsec\,$\times$\,0.96\arcsec. \\ 
\noindent
To provide high resolution images of the optical structure of NGC~1614, we obtained pipeline-reduced \textit{Hubble Space 
Telescope} images taken with the Advanced Camera for Surveys in the F435W and F814W filters (see Fig.\,\ref{fig:overlay_co2-1_co1-0_HST}). 
These data were obtained as part of program G0-10592 (A. Evans, P.~I.) and have an angular resolution of 0.05~arcsec~pixel$^{-1}$ 
corresponding to a linear scale of $\sim$15~pc~pixel$^{-1}$.


\section{Results} \label{sec:results}

\subsection{1.3~mm continuum} \label{subsec:continuum}

The 1.3~mm continuum map (Fig.\,\ref{fig:sma_cont+12co2-1}, upper left panel) shows centrally peaked emission with a flux density of 
$\sim$19\,$\pm$\,6~mJy, which agrees well with the values for the 1.3~mm continuum obtained by \citet{wil08}, who observed a total of 
21\,$\pm$\,3~mJy was observed. Our 1.3~mm continuum map appears to be comprised of two components and has a box-like shape that is 
consistent with the structure \citet{wil08} observed in their 1.3~mm continuum map.

\subsection{$^{\rm 12}$CO and $^{\rm 13}$CO\,(2$-$1)} \label{subsec:co2-1}

\begin{table*}[t] 
\begin{minipage}[!h]{\textwidth}
\centering
\renewcommand{\footnoterule}{}
\caption{\small
 Properties of the observed emission in \object{NGC~1614}.}
\label{tab:co_results}
\tabcolsep0.1cm
\begin{tabular}{lcccccc}
\noalign{\smallskip}
\hline
\noalign{\smallskip}
\hline
\noalign{\smallskip}
 & & \multicolumn{2}{c}{integrated flux} & \multicolumn{3}{c}{mass \footnote{We adopted a conversion factor of 
3\,$\times$\,10$^{\rm 20}$~cm$^{\rm -2}$\,(K\,km\,s$^{\rm -1}$)$^{\rm -1}$}}\\
\noalign{\smallskip}
 & & $^{\rm 12}$CO\,(2$-$1) & $^{\rm 13}$CO\,(2$-$1) & $M$\,($^{\rm 12}$CO\,2$-$1) & $M$\,($^{\rm 13}$CO\,2$-$1) & $M$\,(1.3~mm continuum)\\
\noalign{\smallskip}
 & & [Jy\,km\,s$^{\rm -1}$] & [Jy\,km\,s$^{\rm -1}$] & [M$_{\sun}$] & [M$_{\sun}$] & [M$_{\sun}$] \\
\noalign{\smallskip}
\hline
\noalign{\smallskip}
Whole system         & & 203.3\,$\pm$\,9.8 & $<$ 7.7 & 2.12\,$\times$\,10$^{\rm 9}$ & $<$ 2.74\,$\times$\,10$^{\rm 5}$ & 1.03\,$\times$\,10$^{\rm 9}$ \\
Ring                 & & 65.4\,$\pm$\,6.9  & $<$ 1.0 & 8.29\,$\times$\,10$^{\rm 8}$ & $<$ 3.25\,$\times$\,10$^{\rm 4}$ & 3.94\,$\times$\,10$^{\rm 8}$ \\
Dust lane connection & & 25.8\,$\pm$\,5.3  & $<$ 1.0 & 2.54\,$\times$\,10$^{\rm 8}$ & $<$ 5.40\,$\times$\,10$^{\rm 4}$ & 2.33\,$\times$\,10$^{\rm 7}$ \\
Northern extension   & & 22.7\,$\pm$\,6.2  & $<$ 3.0 & 3.32\,$\times$\,10$^{\rm 8}$ & $<$ 1.39\,$\times$\,10$^{\rm 5}$ & 1.25\,$\times$\,10$^{\rm 8}$ \\
%
%
\noalign{\smallskip}
\hline
\end{tabular}
\end{minipage}
\end{table*}

\begin{figure}[ht]
  \centering
    \includegraphics[width=0.4\textwidth,angle=-90]{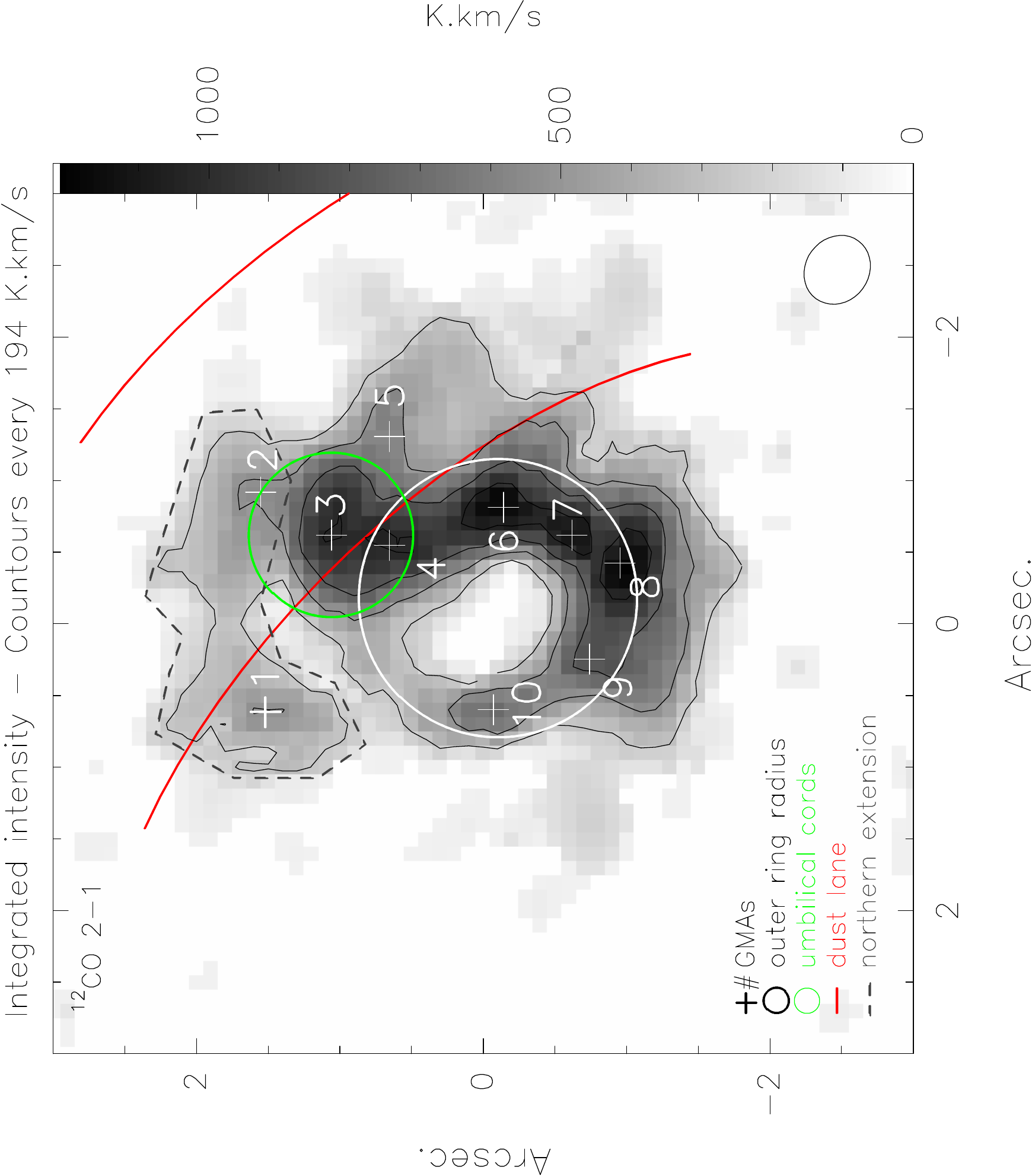}
  \caption{\footnotesize Description of important geometrical structures in the integrated intensity map of the CO\,(2$-$1) emission in 
  \object{NGC~1614}. The extent and position of the starburst ring, the \textit{northern extension}, the umbilical cords (connection to 
  the dust lane) to the north and west of the ring that contain GMAs 2, 3, 4, and 5 and the GMAs are indicated. To clearly distinguish 
  between the GMAs and their positions, their respective numbers are indicated as well.}
  \label{fig:12co2-1_annotations}
\end{figure}

\subsubsection{CO\,(2$-$1) integrated intensity} \label{subsubsec:integr_intensity}

The upper right panel of Fig.\,\ref{fig:sma_cont+12co2-1} shows the $^{\rm 12}$CO\,(2$-$1) integrated intensity map derived from 
our data. The $^{\rm 13}$CO\,(2$-$1) line was not detected at 3$\sigma$ a level of 17.26~Jy\,km\,s$^{\rm -1}$ in these SMA 
observations, which agrees with values from \citet{wil08}, although they detected the line. The higher spatial resolution of 
our Vext data might lead to a non-detection of lower density $^{\rm 13}$CO\,(2$-$1) gas that \citet{wil08} were able to find with 
their lower resolution data. The $^{\rm 12}$CO emission appears as a ring-like structure without emission at its center, with a 
gas component, that might connect to the dust lane (hereafter \textit{umbilical cords}), to the northwest and an extension of 
lower surface brightness gas to the north (see Fig.\,\ref{fig:12co2-1_annotations}). Both within and slightly beyond this ring 
structure hot spots of $^{\rm 12}$CO are clearly identifiable. These are discussed in Sect.\,\ref{subsec:GMAs}.\\
\indent
The highest flux density is 203~Jy\,km\,s$^{\rm -1}$ (Table\,\ref{tab:co_results}). A comparison to previous observations of 
\citet{wil08} shows us that we recovered only about $\sim$30\% of the total flux density. We find a clear asymmetry in the east-west 
gas distribution of the ring-like structure, or possibly tightly wound spiral arms. The eastern side has much less gas than the western 
part. The intensity ratio is I(west)/I(east) $\sim$ 2.7.\\
\indent
Using a CO-to-H$_{\rm 2}$ conversion factor of 
$X_{\rm CO}$\,=\,3\,$\times$\,10$^{\rm 20}$ cm$^{\rm -2}$\,(K\,km\,s$^{\rm -1}$)$^{\rm -1}$ \citep{nar11} results in a molecular 
mass of 2.1\,$\times$\,10$^{\rm 9}$~M$_{\sun}$ for the whole system. If we consider the \textit{umbilical cords} 
(Fig.\,\ref{fig:12co2-1_annotations}) as a separate mass entity, the ring itself has a gas mass of 8.3\,$\times$\,10$^{\rm 8}$~M$_{\sun}$ 
(Table\,\ref{tab:co_results}).

\begin{figure*}[ht]
  \begin{minipage}[hbt]{0.5\textwidth}
  \centering
    \includegraphics[width=0.7\textwidth]{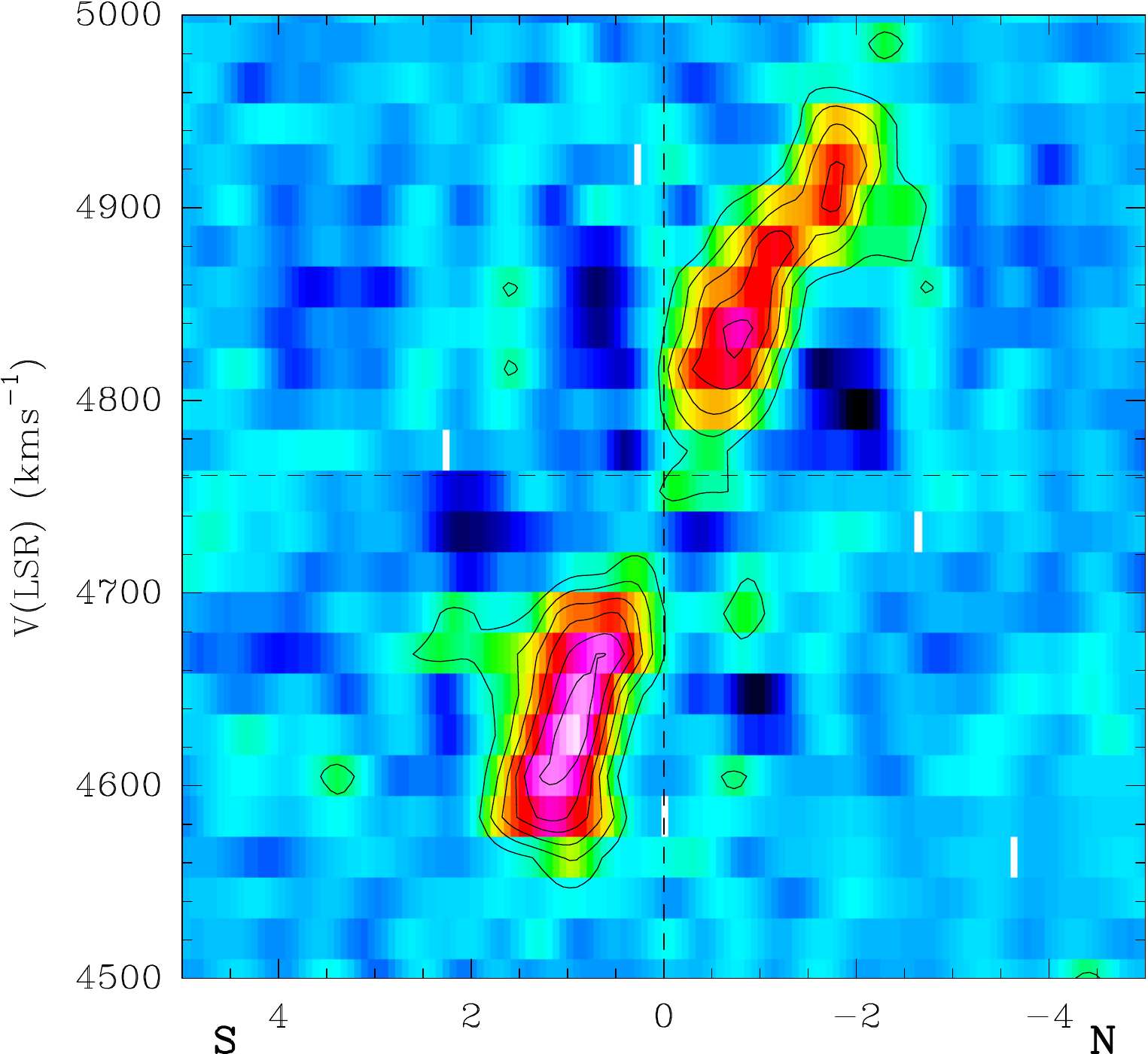}
  \end{minipage}
  \begin{minipage}[hbt]{0.5\textwidth}
  \centering
    \includegraphics[width=0.7\textwidth]{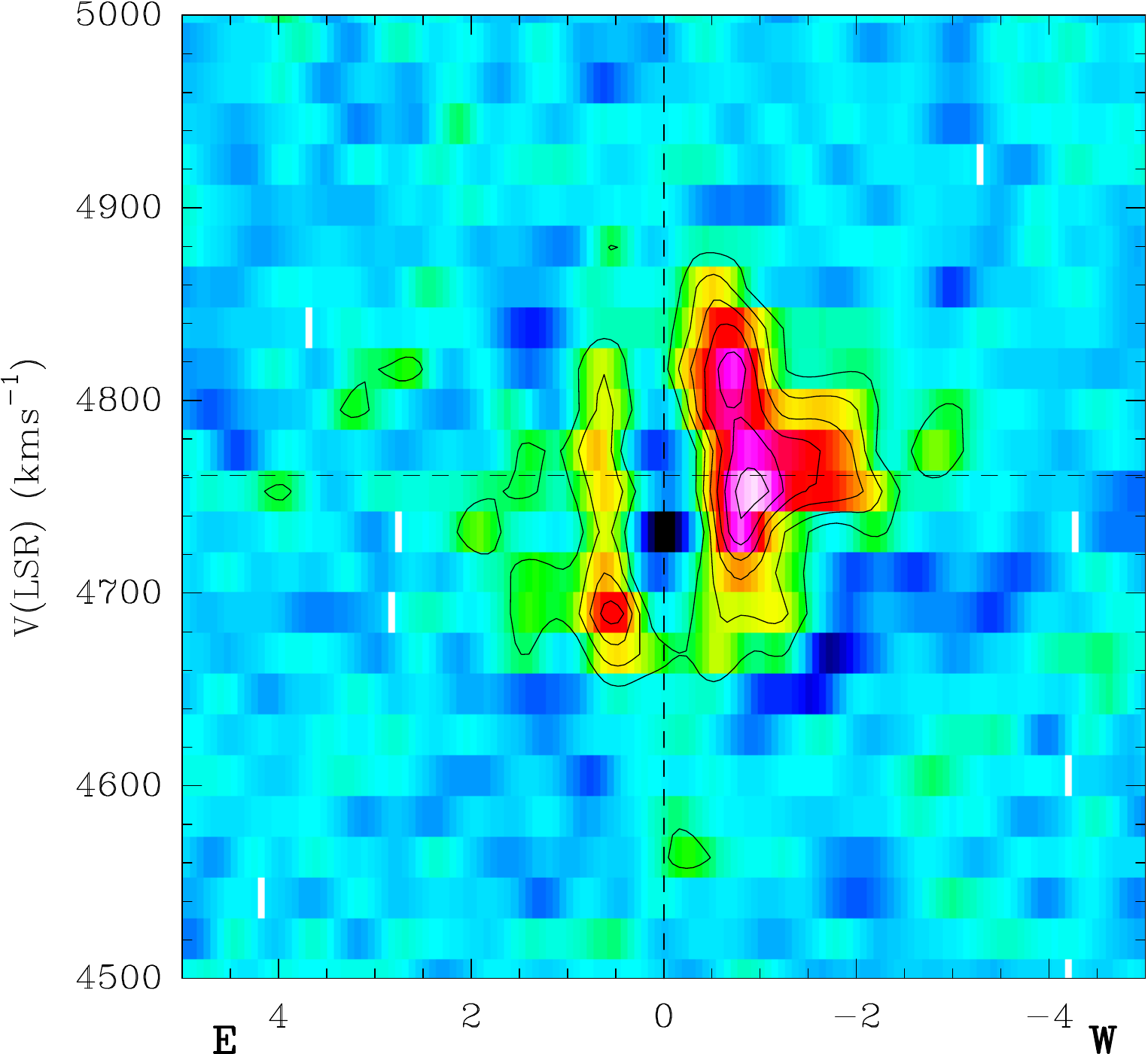}
  \end{minipage}
  \caption{\footnotesize $^{\rm 12}$CO\,(2$-$1) position-velocity diagrams along the major (north-south, \textit{left}) and the 
minor (east-west, \textit{right}) axes of \object{NGC~1614}. The contour levels start at 3$\sigma$ in steps of 3$\sigma$. 
The colors range from 0.1 to 0.28~mJy/beam for the major axis pv diagram and from 0.08 to 0.24~mJy/beam if the minor axis pv 
diagram.}
  \label{fig:12co2-1_pv}
\end{figure*}

\subsubsection{Missing flux and the structure of the CO emission} \label{subsubsec:missing_flux}

In our CO\,(2$-$1) integrated intensity map we recover only 30\% of the CO\,(2$-$1) flux density of the low-resolution SMA observations 
of \citet{wil08}, thus we most likely only see the most compact structures. A comparison of their lower resolution CO\,(2$-$1) 
map and the structures of the high- and low-resolution CO\,(1$-$0) observed by \citet{ols10} shows remarkable similarities in the 
morphology. Hence we suggest that the bulk of the missing flux is located in the northern crossing dust lane (see also 
Fig.\,\ref{fig:overlay_co2-1_co1-0_HST}) - but also in structures that trace the beginning of the southern arm and emission to the 
east of the ring.

\subsubsection{CO\,(2$-$1) velocity field, velocity dispersion, and position-velocity diagrams} 
							\label{subsubsec:pv_velocity_field_dispersion}

The velocity field, with velocities ranging from $\sim$\,4520.42~km\,s$^{\rm -1}$ to 4985.31~km\,s$^{\rm -1}$, is shown in 
Fig.\,\ref{fig:sma_cont+12co2-1} (bottom left panel). It shows a behavior typical for solid-body rotation, for which the probable 
connection to the dust lane follows the behavior of the gas in the ring.\\
\indent
The velocity dispersion distribution of \object{NGC~1614} shows several peaks (Fig.\,\ref{fig:sma_cont+12co2-1}, bottom right 
panel). At these peaks the velocity dispersions rise to 43-51~km\,s$^{\rm -1}$ compared to the underlying material, which 
only shows dispersion values between $\sim$15 and $\sim$33~km\,s$^{\rm -1}$. The velocity dispersion might be biased by missing 
extended structures in our map. This probably does not affect the dispersion of the ring itself because it is compact.\\
\indent
In Fig.\,\ref{fig:12co2-1_pv} the position velocity diagrams along the major (north to south) and minor (east to west) axes 
are shown. They were obtained by averaging over a 1\arcsec -wide slit positioned along the major and minor axes centered on 
the center of the ring. The pattern for solid-body rotation seen in the velocity field is also reflected in the major axis position 
velocity diagram. In the minor axis position velocity diagram we clearly see a difference between the eastern and the western 
part of the ring, with the majority of the gas residing in the western part of the ring.

\subsubsection{$^{\rm 12}$CO\,/\,$^{\rm 13}$CO\,(2$-$1) ratio} \label{subsubsec:13co_12co_ratio}

To estimate the $^{\rm 12}$CO\,/\,$^{\rm 13}$CO\,(2$-$1) ratio we used the integrated intensity maps of 
$^{\rm 12}$CO\,(2$-$1) and $^{\rm 13}$CO\,(2$-$1) integrated over the same velocity range and determined the average flux densities 
(flux density for $^{\rm 13}$CO from the 3$\sigma$ rms noise value) within the same polygon for both transitions. The resulting lower 
limit for the flux density ratio is $^{\rm 12}$CO\,/\,$^{\rm 13}$CO\,(2$-$1) $>$ 15. An investigation of the spectra of \citet{wil08} 
showed that the $^{\rm 12}$CO\,/\,$^{\rm 13}$CO\,(2$-$1) line ratio for the peak (central) flux density is $\sim$ 22, while the ratio for 
the integrated flux density (in the whole map) is $\sim$ 38. One can see from their data that the $^{\rm 13}$CO\,(2$-$1) flux density 
hardly rises going from peak flux densities to integrated intensities, while the $^{\rm 12}$CO\,(2$-$1) flux densities increase by more 
than a factor of 2. From this it seems that the $^{\rm 12}$CO\,/\,$^{\rm 13}$CO\,(2$-$1) ratio is increasing with radius in NGC~1614, a 
notion that should be confirmed at higher sensitivity and resolution; this would be reminiscent of the situation in the Medusa minor 
merger \citep[\object{NGC~4194},][]{aalto10} and the molecular ring in \object{NGC~6946} \citep{mei04}. This is discussed again briefly 
in Sect.\,\ref{subsubsec:gma_formation}.


\begin{table}[ht]
\begin{minipage}[!h]{0.5\textwidth}
\centering
\renewcommand{\footnoterule}{}
\caption{\small
 Properties of the molecular ring in NGC~1614.}
\label{tab:ring_properties}
\tabcolsep0.1cm
\begin{tabular}{lc}
\noalign{\smallskip}
\hline
\noalign{\smallskip}
\hline
\noalign{\smallskip}
Property & \\
\noalign{\smallskip}
\hline
\noalign{\smallskip}
Ring center coordinates        & \vspace{0.1cm} RA: 04:34:00.036, DEC: -08:34:45.08 \\
Molecular mass $M$$_{\rm mol}$ & 8.29\,$\times$\,10$^{\rm 8}$~M$_{\sun}$ \\
Dynamical mass $M$$_{\rm dyn}$ & 1.85\,$\times$\,10$^{\rm 9}$~M$_{\sun}$ \\
West-east intensity ratio      & 2.7\\
Radius 			       & 0.75\arcsec\ ($\sim$ 231~pc) \\
GMAs on the ring  	       & 6 \\
GMAs off the ring  	       & 4 \\
\noalign{\smallskip}
\hline
\end{tabular}
\end{minipage}
\end{table}

\subsection{Properties of the molecular ring} \label{subsec:ring_properties}

\begin{figure*}[ht]
  \hspace{-1.0cm}
  \begin{minipage}[hbt]{0.5\textwidth}
  \centering
    \includegraphics[width=1.225\textwidth]{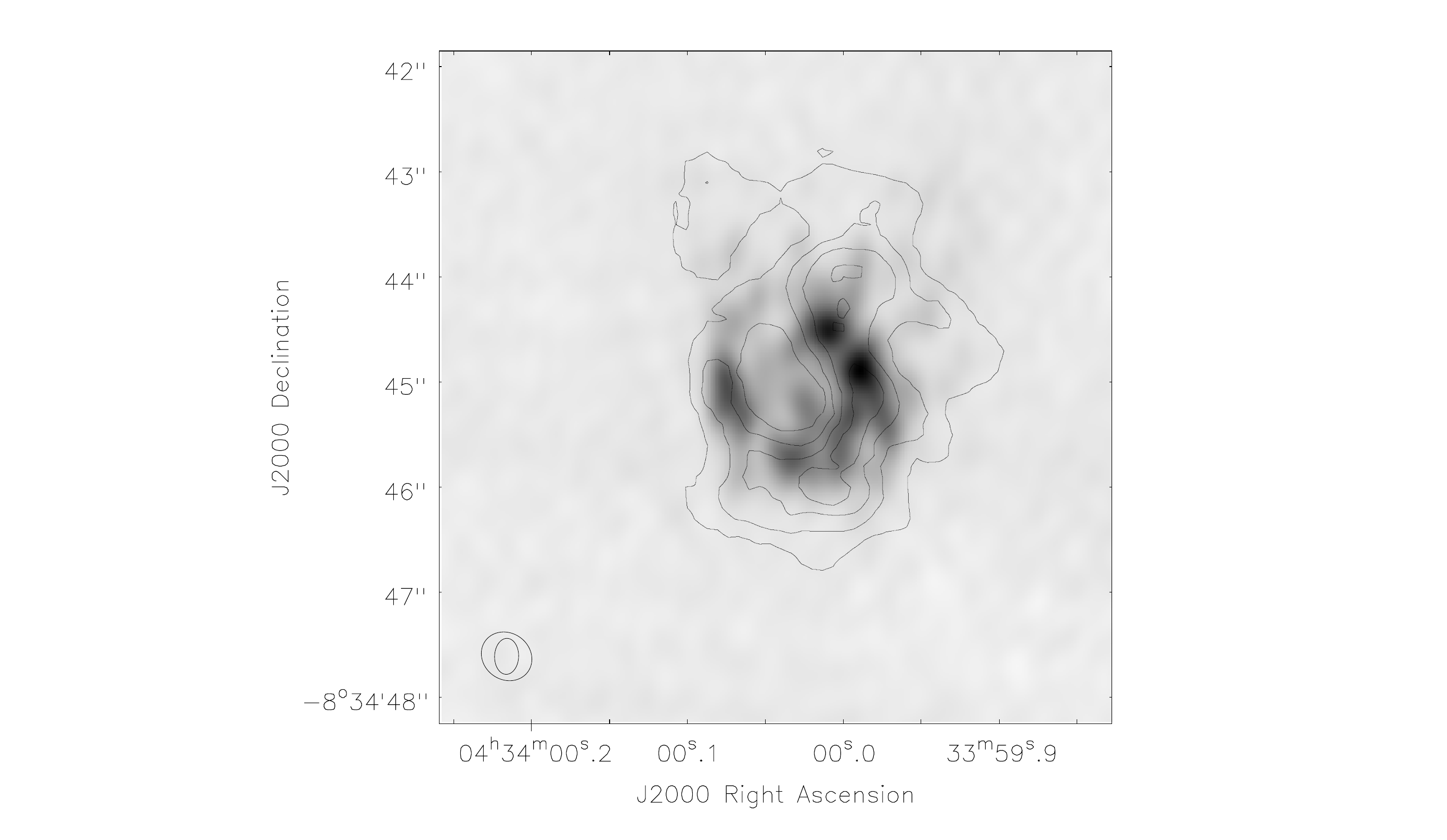}
  \end{minipage}
  \begin{minipage}[hbt]{0.5\textwidth}
  \hspace{-2.1cm}
  \centering
    \includegraphics[width=1.225\textwidth]{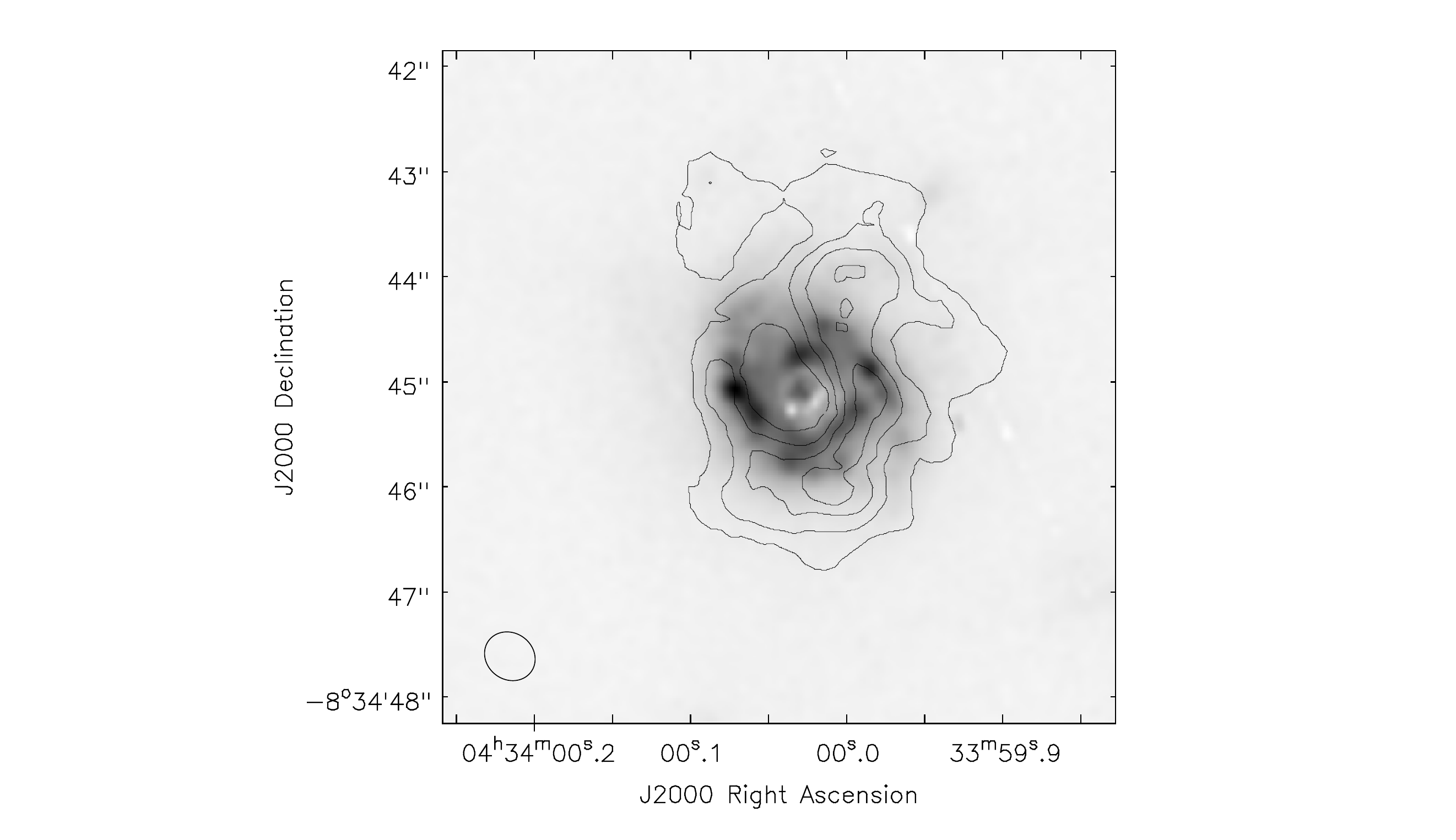}
  \end{minipage}
  \caption{\footnotesize Overlay of the CO\,(2$-$1) integrated intensity contours on the VLA 8~GHz emission \citep[grayscale, 
  \textit{left},][]{ols10} and overlay of the CO\,(2$-$1) integrated intensity emission (contours) on the Pa$\alpha$ emission 
  \citep[grayscale, \textit{right},][]{ols10}. The contour levels for the CO\,(2$-$1) emission are at 18.5, 41, 60.5, and 81\% of the 
  emission peak (same levels as in Fig.\,\ref{fig:sma_cont+12co2-1}).} 
 \label{fig:overlay_co2-1_vla_paalpha}
\end{figure*}

Figs.\,\ref{fig:sma_cont+12co2-1}, \ref{fig:12co2-1_annotations}, \ref{fig:overlay_co2-1_vla_paalpha}, and \ref{fig:annular_profiles} 
show that we observe a ring-like structure in \object{NGC~1614} with additional molecular gas emission to the north, the northeast, and 
the west. From the inside out, the structure evolves via \textit{umbilical cords}, a gas component that possibly connects the ring to 
the dust lane located to the northeast, and then into a \textit{northern extension} of lower surface density material (see 
Fig.\,\ref{fig:12co2-1_annotations}). In contrast to the observations from \citet{wil08} and \citet{ols10}, we clearly filtered out the 
more extended gas in this region. The angular resolution of our observations allowed us to resolve the width of the ring. In contrast to 
studies of NGC~1614 at other wavelengths, such as radio continuum and Pa$\alpha$, these $^{\rm 12}$CO\,(2$-$1) observations detected no 
emission at the center of the ring (Fig.\,\ref{fig:overlay_co2-1_vla_paalpha}).\\
\indent
To obtain a better overview of the properties of the different components in this object, we also determined the masses for components 
outside the ring. We assumed a radius of the ring of 0.75\arcsec\ ($\sim$231~pc) and a width of 0.5\arcsec. On this basis the mass for 
the ring is $M$(H$_{\rm 2}$)\,=\,8.3\,$\times$\,10$^{\rm 8}$~M$_{\sun}$ (Table\,\ref{tab:ring_properties}). The \textit{umbilical cords} 
to the northeast have a mass of $M$(H$_{\rm 2}$)\,=\,2.5\,$\times$\,10$^{\rm 8}$~M$_{\sun}$ and the mass of the \textit{northern 
extension} is $M$(H$_{\rm 2}$)\,=\,3.3\,$\times$\,10$^{\rm 8}$~M$_{\sun}$. We can now also calculate the dynamical mass for the ring by 
considering the radius and the width of the ring, as well as a rotational velocity of 187~km\,s$^{\rm -1}$, estimated from the position 
velocity diagram. Because the angle of the ring to the plane of the sky is unknown, an uncertainty factor of $\cos(\theta)$ has to be 
considered. Taking this into account, we determined the dynamical mass of the ring to be 
$M$$_{\rm dyn}$\,=\,1.85\,$\times$\,10$^{\rm 9}$\,$\cos(\theta)$~M$_{\sun}$, which is a factor of $\sim$2 higher than the molecular mass 
of the ring. The viewing angle to the molecular ring seems to be face-on, which justifies a $\cos(\theta)$ of 1. A low inclination 
would not change the result drastically.

\begin{table*}[ht]
\begin{minipage}[!h]{\textwidth}
\centering
\renewcommand{\footnoterule}{}
\caption{\small
 Properties of the identified GMAs.}
\label{tab:GMA_properties}
\tabcolsep0.1cm
\begin{tabular}{lcccccccc}
\noalign{\smallskip}
\hline
\noalign{\smallskip}
\hline
\noalign{\smallskip}
GMA &  & RA(2000) & DEC(2000) & \multicolumn{2}{c}{   size (diam.)\footnote{The source size was determined by fitting an elliptical 
Gaussian to the position of the respective GMA in the uv table. The values given here represent the major and minor axis values resulting 
from the fitting procedure. The typical (average) error for the fit to the GMA sizes is about 20\%. }} & 
$v$$_{\rm 0}$ & dispersion & mass \\
\noalign{\smallskip}
        & & [h] [m] [s] & [\degr] [\arcmin] [\arcsec] & [\arcsec] & [pc] & [km\,s$^{\rm -1}$] & [km\,s$^{\rm -1}$] & [M$_{\sun}$] \\
\noalign{\smallskip}
\hline
\noalign{\smallskip}
\underline{\textit{GMAs in the ring:}} & &  &   &                  & &              &       &                              \\
GMA\,4 & & 04:34:00.004 & --08:34:44.37 & 0.6\,$\times$\,1.0 & 250\,$\times$\,336 & 4813\,$\pm$\,7 & 61\,$\pm$\,6 & 5.70\,$\times$\,10$^{\rm 7}$\\
GMA\,6 & & 04:33:59.985 & --08:34:45.08 & 0.6\,$\times$\,0.9 & 195\,$\times$\,268 & 4737\,$\pm$\,5 & 64\,$\pm$\,4 & 9.74\,$\times$\,10$^{\rm 7}$\\
GMA\,7 & & 04:33:59.998 & --08:34:45.57 & 0.8\,$\times$\,1.1 & 261\,$\times$\,341 & 4670\,$\pm$\,7 & 38\,$\pm$\,4 & 5.97\,$\times$\,10$^{\rm 7}$\\
GMA\,8 & & 04:34:00.012 & --08:34:45.97 & 1.3\,$\times$\,0.8 & 390\,$\times$\,260 & 4609\,$\pm$\,6 & 73\,$\pm$\,5 & 9.36\,$\times$\,10$^{\rm 7}$\\
GMA\,9 & & 04:34:00.067 & --08:34:45.56 & 0.6\,$\times$\,1.2 & 192\,$\times$\,368 & 4671\,$\pm$\,6 & 54\,$\pm$\,5 & 5.98\,$\times$\,10$^{\rm 7}$\\
GMA\,10 & & 04:34:00.082 & --08:34:45.01 & 0.5\,$\times$\,0.7 & 168\,$\times$\,204 & 4756\,$\pm$\,7 & 40\,$\pm$\,5 & 4.96\,$\times$\,10$^{\rm 7}$\\
\noalign{\smallskip}
\underline{\textit{GMAs outside the ring:}} & & &   &                  & &              &       &                              \\
GMA\,1 & & 04:34:00.087 & --08:34:43.45 & 0.6\,$\times$\,1.1 & 197\,$\times$\,326 & 4919\,$\pm$\,7 & 57\,$\pm$\,7 & 8.29\,$\times$\,10$^{\rm 7}$\\
GMA\,2 & & 04:33:59.989 & --08:34:43.38 & 1.1\,$\times$\,0.5 & 336\,$\times$\,157 & 4901\,$\pm$\,7 & 31\,$\pm$\,4 & 1.64\,$\times$\,10$^{\rm 7}$\\
GMA\,3 & & 04:33:59.998 & --08:34:43.97 & 0.8\,$\times$\,1.1 & 250\,$\times$\,336 & 4875\,$\pm$\,5 & 53\,$\pm$\,3 & 6.23\,$\times$\,10$^{\rm 7}$\\
GMA\,5 & & 04:33:59.946 & --08:34:44.37 & 1.8\,$\times$\,0.6 & 566\,$\times$\,180 & 4814\,$\pm$\,7 & 48\,$\pm$\,5 & 3.70\,$\times$\,10$^{\rm 7}$\\
\noalign{\smallskip}
\hline
\end{tabular}
\end{minipage}
\end{table*}

\subsection{Giant molecular associations (GMAs) in \object{NGC~1614}} \label{subsec:GMAs}

We found ten clumps in (GMA\,4, GMA\,6, GMA\,7, GMA\,8, GMA\,9, GMA\,10) and outside (GMA\,1, GMA\,2, GMA\,3, GMA\,5) the structure of 
the molecular ring (Fig.\,\ref{fig:12co2-1_annotations}). They have masses between 
$M$(H$_{\rm 2}$)\,=\,1.6\,$\times$\,10$^{\rm 7}$~M$_{\sun}$ and $M$(H$_{\rm 2}$)\,=\,9.4\,$\times$\,10$^{\rm 7}$~M$_{\sun}$ 
(Table\,\ref{tab:GMA_properties}), which places these objects in the mass range for GMAs, so-called giant molecular associations 
\citep{vogel88}.\\
\indent
The velocity dispersion map (Fig.\,\ref{fig:sma_cont+12co2-1}, bottom right panel) shows a rise in line widths at the positions of all 
GMAs in the ring (GMA\,4, GMA\,6, GMA\,7, GMA\,8, GMA\,9). The dispersion values range between 31.1~km\,s$^{\rm -1}$ for GMA\,2 and 
72.9~km\,s$^{\rm -1}$ for GMA\,8. The GMAs outside the ring have an average dispersion of 47.0~km\,s$^{\rm -1}$. The average dispersion 
for the GMAs in the ring is 55.0~km\,s$^{\rm -1}$.\\
\indent
We ran a decomposition of the CO data cube into individual structures, using the task Gaussclumps 
\citep[developed initially by][]{stu90} in GILDAS/Mapping. The algorithm searches iteratively for 3D Gaussians in the data cube until 
it reaches a specified threshold value. To derive reliable results, we set this threshold to a value of 6$\sigma$. Under these 
conditions, the data cube can be decomposed into ten different Gaussian clumps, whose properties are listed in 
Table\,\ref{tab:GMA_properties} while the positions are indicated in Fig.\,\ref{fig:12co2-1_annotations}. Four of the five brightest 
clumps (GMA\,4, GMA\,6, GMA\,8, and GMA\,9), which we call GMAs, are clearly associated with the ring, where the CO emission is the 
brightest. The one GMA that is very bright (GMA\,3) but does not lie in the ring is situated at the point where the molecular gas in 
the ring is connected to the dust lanes via the feeding lines (\textit{umbilical cords}, see Sect.\,\ref{subsubsec:feeding} and 
\ref{subsubsec:gma_formation}). The residual map is dominated by fainter emission located to the north of the ring. The resulting 
source sizes (Table\,\ref{tab:GMA_properties}) range between 168~pc\,$\times$\,204~pc and 390~pc\,$\times$\,260~pc for GMAs outside the 
molecular ring and 336~pc\,$\times$\,157~pc up to 566~pc\,$\times$\,180~pc for GMAs in the ring. Moreover, the GMAs outside the 
ring are more elliptical in shape, i.e., the major-to-minor axis ratio is much higher than for the GMAs in the ring, especially for the 
GMAs outside the ring that are situated in the dust lanes. Assuming spherical morphologies (for comparison purposes only), the average 
diameter of a GMA outside the ring is 273\,$\pm$\,20~pc, whereas the GMAs in the ring show an average diameter of 259\,$\pm$\,24~pc.


\section{Discussion} \label{sec:discussion}

\subsection{Origin of the ring} \label{subsec:origin}

\subsubsection{``Feeding'' of the ring: The connection to the dust lanes} \label{subsubsec:feeding}

\citet{bour05} found that in minor mergers gas brought in by a disturbing galaxy companion is generally found at large radii in the 
merger remnant. The gas returns to the system via tidal tails and often forms rings - polar, inclined or equatorial - that will appear as 
dust lanes seen edge-on. In the CO\,(2$-$1) gas distribution of \object{NGC~1614} (Fig.\,\ref{fig:sma_cont+12co2-1}) we found that the 
feature labelled \textit{umbilical cords} appears to be associated with a dust lane northwest of the nucleus of NGC~1614, which has 
previously been identified with the CO\,(1$-$0) emission by 
\citet[][see also Fig.\,\ref{fig:overlay_co2-1_co1-0_HST} in this paper]{ols10}. This link to the dust lane might be the connection point 
that fuels the starburst ring and is comparable with the twin-peak morphology in the rings of barred spiral galaxies (e.g. 
object{NGC~1097}, \object{NGC~1365}, see Sect.\,\ref{subsec:comparison}). However, for NGC~1614 we only found one connection point, and 
a possible a stellar bar and its properties is still under debate \citep[see for example the discussions in][]{cha99,ols10}. This 
connection could be due to orbit crowding of inflowing gas streams (\textit{umbilical cords}, traced by the CO\,(1$-$0) and the dust lane 
to the northwest in Fig.\,\ref{fig:overlay_co2-1_co1-0_HST}) at the branch-off of the ring, where gas encounters shocks and then migrates 
to new orbits where it accumulates in the shape of the starburst ring. The gas at the western and southern edges of the ring is 
consistent with an additional dust lane component that crosses the ring from the west to the southeast (see 
Fig.\,\ref{fig:overlay_co2-1_co1-0_HST}). This additional component is possibly the site of a second connection between the starburst 
ring and the dust lanes in this system, but this remains to be confirmed.\\
\indent
As observed in high-resolution CO\,(1$-$0) \citep{ols10}, the starburst ring has the same velocity structure as the underlying galaxy, 
which we do not see because the lower surface density gas is filtered out in our high-resolution observations. Within the ring the 
rotation is higher than in the rest of the galaxy. The gas slows down relative to the galaxy's rotation curve toward the edges of the 
CO\,(2$-$1) distribution at the connection point, where the dust lane connects to the ring in the northern part. The gas distribution in the ring may be lopsided because the whole ring is fueled/fed from one side only, which might also be the reason for the asymmetry of the ring itself.\\
\indent
We found that the CO\,(2$-$1) emission resolved in our interferometric observations is located at the unresolved center of the larger 
scale molecular gas reservoir (CO\,(1$-$0)). However, with the present data sets we were unable to verify possible streaming motions 
between the larger scale and smaller scale molecular gas reservoirs. Therefore we are unable to determine the inflow rate of the gas. To 
correct this, higher resolution CO\,(1$-$0) observations are needed.\\ 
\indent 
With the large reservoir of available gas in the ring and the star formation rate (SFR) of $\sim$5.2~M$_{\sun}$\,yr$^{\rm -1}$ determined 
from the $^{\rm 12}$CO\,(2$-$1) emission, taking the gas mass in the ring and its surface density into account, the star formation in the 
center of NGC~1614 could go on for another $\sim$1.6\,$\times$\,10$^{\rm 8}$~yr unless processes/events occur that stop the gas transport 
via the feeding lines.

\begin{figure*}[t]
 \sidecaption
  \centering
    \includegraphics[width=11cm]{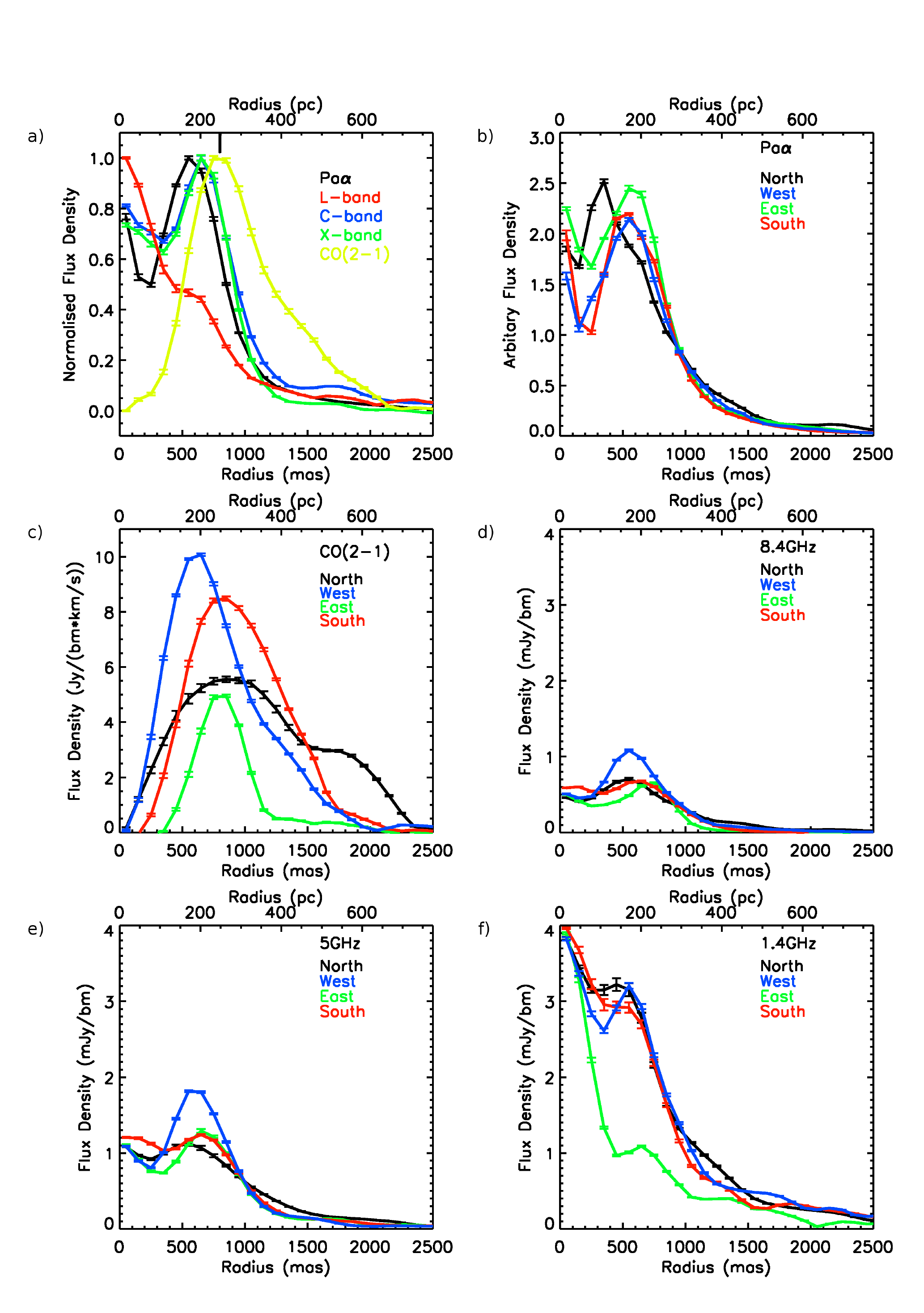}
  \caption{\footnotesize Comparison of the annular profiles of the new CO\,(2$-$1) data with Pa$\alpha$, 8.4~GHz ($X$-band), 
  5~GHz ($C$-band) and 1.4~GHz ($L$-band) data from \citealt{ols10} (a), \textit{top left}). These profiles, obtained by azimuthally
  averaging the flux density in 0.1\arcsec\ wide annular circular rings, are nominally centered on the center of the Pa$\alpha$ ring. 
  Profiles of each tracer in different regions of the ring are presented from \textit{top right} to \textit{bottom left}: Pa$\alpha$ (b), 
  CO\,(2$-$1) (c\,), 8.4~GHz (d), 5~GHz (e) and 1.4~GHz (f). Spatial resolutions of the data used for the profiles are $\sim$0.2\arcsec\ 
  for Pa$\alpha$, $\sim$0.3\arcsec\ for the $L$-, $C$- and $X$-band observations and the spatial resolution for the CO data is 
  $\sim$0.5\arcsec.}
  \label{fig:annular_profiles}
\end{figure*}

\subsubsection{Formation of GMAs in the ring} \label{subsubsec:gma_formation}

In Sect.\,\ref{subsubsec:feeding} we established that there is a connection between the dust lanes in \object{NGC~1614} and the 
molecular ring seen in $^{\rm 12}$CO\,2$-$1 emission. Prominent features of the ring and the surrounding gaseous material are ten GMAs. 
In the prominent minor axis dust lane north of the ring, however, we detect no GMAs - although this is the main location of the 
bulk of the CO\,(2$-$1) flux \citep{wil08}. Hence, the CO\,(2$-$1) emission here is either emerging from smaller molecular clouds with 
a surface brightness too low to be detected by us - or alternatively the molecular gas is in the form of diffuse, unbound molecular 
clouds that are filtered out by the lack of short-spacing baselines in our data. The latter scenario would be consistent with the 
increase in $^{\rm 12}$CO/$^{\rm 13}$CO\,(2$-$1) ratio with radius reported in Sect.\,\ref{subsubsec:13co_12co_ratio}. This scenario may 
have similarities to that found for \object{NGC~6946} \citet{mei04}, where molecular gas outside the central ring was found to be of 
lower density and more diffuse than the dense, self-gravitating gas in the ring. This is reflected as a rapid change in the interstellar 
medium (ISM) from dense star-forming cores in a central ring to diffuse, low-density molecular gas in and behind the molecular arms. This 
results in $^{\rm 12}$CO/$^{\rm 13}$CO\,(2$-$1) intensity ratios in the central ring that are lower than in the diffuse gas on larger 
scales - which also seems to be the case for NGC~1614 (see Sect.\,\ref{subsubsec:13co_12co_ratio}). Therefore, the absence of GMAs 
outside the ring leeds us to conclude that the GMAs in NGC~1614 are formed as a result of some form of orbit-crowding within the nuclear 
ring from gas funneled from the dust lane reservoir.\\
\indent
Analogies for the GMA formation processes could be \object{M~51} \citep{koda09} or \object{IC~342} \citep[][see also 
Sect.\,\ref{subsubsec:comparison_rings_gmas} for a more detailed comparison with other GMA-bearing galaxies]{hiro11}. In \object{M~51} 
the GMAs are only found in the spiral arms. They are formed upon entering the spiral arms through coagulation triggered by spiral-arm 
streaming motions and are disrupted again by fragmentation upon leaving the spiral arms.\\
\indent
For NGC~1614 we propose the following scenario: Smaller sized GMCs are funneled along the \textit{umbilical cords} toward the molecular 
ring. At the transition from the dust lane to the molecular ring we assume that crowding processes (e.g. orbital crowding) in the mergers 
take place that lead to the collisional coagulation of individual GMCs. These cloud-cloud collisions could also trigger the onset of star 
formation \citep{sco86,tan00} - the dissipation of excess kinetic energy during the collisions is a prerequisite for star formation 
\citep{hiro11}. Observations at higher angular resolution and sensitivity are necessary to resolve the molecular gas properties of the 
low surface brightness GMCs in the \textit{umbilical cords}.

\subsection{Tracers of star formation}

\subsubsection{Annular profiles} \label{subsubsec:ann_profiles}

\noindent
In Fig.\,\ref{fig:annular_profiles} we present the annular profiles of Pa$\alpha$, CO\,(2$-$1), radio $L$- (1.4~GHz), $C$- (5~GHz), and 
$X$ (8.4~GHz)-band. The averaged annular profiles for the \object{NGC~1614} ring are shown for all tracers 
(Fig.\,\ref{fig:annular_profiles} upper left panel). We simply azimuthally averaged the flux density in a series of 0.1\arcsec -wide 
annular circular rings (i.e., we excluded inclination effects). We did not exclude any parts of the annulus that might be atypical in 
any band. We also set the ring center in the middle of the Pa$\alpha$ where there is a slight central feature and used this position in all bands. This matches the weak $C$- and $X$-band central feature quite well.\\ 
\indent
We also divided the ring into four sections, north (N), east (E), west (W) and south (S) and plotted semi-annuli to study any positional 
shift in the relative radii of all five tracers (Fig.\,\ref{fig:annular_profiles}).\\
\indent
For the average annuli we see that the $L$-band profile is the tracer with the strongest relative nuclear peak. The ring (or ring-like 
structure) is visible in $L$-band (indicated by the step in the profile) but not quite as pronounced and possibly not with exactly the 
same central position. As expected, the radio $C$- and $X$-band data trace each other very well - and the ring radius may be shifted by 
0.1\arcsec\ compared to that of Pa$\alpha$ - but the difference is very small and within errors. The average CO\,(2$-$1) profile is 
shifted outwards by 0.2\arcsec\ from that of Pa$\alpha$. The nuclear region is obviously void of any CO\,(2$-$1) emission 
(that can be recovered in Vext by the SMA).\\
\indent
However, when inspecting the N, E, W, and S cuts, we find that there are large directional differences in the peak radii of the different 
tracers. All W cuts show a peak radius of 0.6\arcsec\ for all tracers - the agreement is almost perfect and there is a well-defined 
local maximum for all tracers. In contrast, the strongest discrepancy is seen in the N cut between Pa$\alpha$ and CO\,(2$-$1). 
Pa$\alpha$ peaks at 0.3\arcsec\ with a sharp profile while CO\,(2$-$1) peaks at 0.9\arcsec\ with a large, rounded, extended (even 
double-peaked) structure. A similar but less pronounced effect can be seen in the S cut.\\
\indent
Part of the explanation for this discrepancy can be seen in Fig.\,\ref{fig:overlay_co2-1_vla_paalpha} which clearly shows the north-south 
elongated distribution of CO\,(2$-$1) with respect to Pa$\alpha$. The correlation between the tracers is good in the west - but poorer 
in the north and south because of the additional contribution to the CO\,(2$-$1) emission that does not exclusively come from the ring itself, but also from the combined CO\,(2$-$1) contributions from the dust lane and its connection point to the ring.\\
\indent
The annuli are discussed in more detail in Sect.\,\ref{subsubsec:comparison_medusa}.

\subsubsection{``Wildfire'' vs. dynamics} \label{subsubsec:wild_vs_dyn}

There have been suggestions for different evolutionary tracks for the development of a starburst ring in \object{NGC~1614}. \citet{alo00} 
suggested a ``wildfire''-like scenario in which the ring was formed by a nuclear starburst that progresses outward and already has 
consumed almost the whole gas in its center. In contrast to this, \citet{ols10} proposed a scenario in which gas is piled up at an inner 
Lindblad resonance to form the ring.\\
\indent
Supporting the wildfire origin of the starburst ring is the fact that the Pa$\alpha$ and the radio continuum emission are distributed 
consistently inside the starburst ring, as seen in CO\,(2$-$1). The same is true for 1.6~$\mu$m emission ($H$-band) observed by 
\citet{haan11}. The emission is centrally peaked like the Pa$\alpha$ and the radio continuum emission. But in contrast to the other 
tracers, the 1.6~$\mu$m emission is distributed in a broken-ring structure that has an opening to the northwest \citep{haan11}. Our 
observations of CO\,(2$-$1) show, however, that the molecular ring is open to the northeast. The radio continuum suggests, furthermore, 
that an older starburst lies in the ring center at the nucleus. An alternative explanation for the 1.4~GHz emission in the center could 
be an additional synchrotron component, possibly from a weak AGN. However, the distance between the tracers varies with the direction, 
which could also be indicative of spatial shifts between the tracers that are not caused by wildfires.\\
\indent
It was also suggested that the starburst ring in NGC~1614 could be caused by an inner Lindblad  resonance \citep{ols10}. One might argue 
that this process only takes place in barred galaxies, and it is still debated whether there is a stellar bar in NGC~1614 or not 
\citep[see e.g.,][who found an oval distortion in their $K$-band data that is consistent with a bar]{ols10}. But galaxies with molecular 
rings have been found that have no bars \citep[e.g. \object{NGC~278} \& \object{NGC~7217},][]{kna04,sar07}. It has been shown that other 
physical properties, e.g., a weak oval distortion \citep{but95}, a dissolved bar \citep{ver95}, tidal effects of a companion galaxy 
\citep{com88b}, or a recent minor merger \citep{kna04,maz06} can cause the same effect as a bar: the non-axisymmetric potential caused by 
the bar can channel gas to the nucleus, which is slowed down near or at the inner Lindblad resonance. \citet{elm94} proposed a scenario 
in which the gravitational collapse of gas that is caught in an inner Lindblad resonance within a starburst ring, fueled by galactic bars 
or strong spiral arms, can lead to the formation of giant clouds with spots of intense star formation. We do see these hot spots in 
Figs.\,\ref{fig:overlay_co2-1_co1-0_HST}, \ref{fig:sma_cont+12co2-1}, and \ref{fig:overlay_co2-1_vla_paalpha}. Another argument against 
the wildfire scenario is the asymmetry we found in the CO\,(2$-$1) emission, whereas the distribution of star formation tracers, such as 
Pa$\alpha$ and radio continuum emission, is distributed very symmetrically. Conversely, Pa$\alpha$ and the radio continuum agree very 
well with each other but not with the CO\,(2$-$1) distribution, which supports the idea of a wildfire because CO traces gas properties 
that occur earlier, i.e., farther outward, in the star formation process than those traced by Pa$\alpha$ and the radio continuum.\\
\indent
An alternative theory that can possibly explain the spread in distribution of the star formation tracers was suggested by \citet{lou01a}, 
\citet{lou01b}, and \citet{lou03}. These authors highlight the importance of density waves in spiral galaxies as triggers for AGN and 
star-bursting events: a bar or companion induces spiral density waves at the inner Lindblad resonance 
\citep[][and references therein]{lou01a}. \citet{lou01b} found that fast magneto-hydrodynamic density waves (FMDWs) play an important 
dynamic role in developing the circumnuclear region of spiral galaxies with bars. The critical parameter seems to be the dampening of the 
FMDWs, which can lead to the onset of AGN activity (weak density wave dampening), to the formation of circumnuclear rings (strong 
dampening) and a mixture of circumnuclear and AGN activity (intermediate dampening) inside the inner Lindblad resonance. The two latter 
processes might play a role in NGC~1614.\\
\indent
In summary, there is no clear tendency toward the wildfire scenario, density waves, or the idea of a Lindblad resonance as a cause for 
the starburst ring in NGC~1614. From our study of the connection between the molecular gas in the ring and the galaxies dust lanes, we 
propose the following. Star formation in the molecular ring occurs at a Lindblad resonance and is replenished with molecular gas via the 
dust lanes. This is a result of orbital crowding or density wave phenomena. The morphology of the ring as seen in Pa$\alpha$ could be 
caused by a wildfire independent of this star formation. We suggest that both processes coexist at the same time, but that the molecular 
ring is responsible for the starburst. To be able to distinguish between the three competing models, several observations might be 
useful: CO\,(1$-$0) observations with a resolution higher than currently available would enable us to study the possible connection 
between the CO\,(1$-$0) and the CO\,(2$-$1) gas reservoirs, CO data with a higher sensitivity would provide insight into the low surface 
brightness gas content of NGC~1614 and would allow us to continue studying the GMA formation, and observations of shock tracers, such as 
SiO, would help tracing a possible outward movement of the star formation by showing the places of shocked gas within the ISM.

\begin{table*}[ht]
\begin{minipage}[!h]{\textwidth}
\centering
\renewcommand{\footnoterule}{}
\caption{\small
 Overview of the parameters of galaxies with molecular rings compared to NGC~1614.}
\label{tab:vgl_gal_mol_rings}
\tabcolsep0.1cm
\begin{tabular}{lcccccccccc}
\noalign{\smallskip}
\hline
\noalign{\smallskip}
\hline
\noalign{\smallskip}
 Galaxy & & $D$ & Hubble & Merger/ & Ring radius & CO central & Ring\footnote{sym = symmetric, asym = asymmetric} & 
Connection to\footnote{A number given denotes the number of connection points of the molecular ring to dust lanes, ``--'' 
indicates that there is no information available \newline \hspace*{2.5mm} on a possible ring-dust lane connection.} & GMAs
\footnote{A number given denotes the number of GMAs found  in the molecular rings and dust lanes, ``--'' indicates that 
there is no information available \newline \hspace*{2.5mm} on GMAs in these objects.} & Reference\footnote{References: 1): 
this work, 2): \citet{hsieh08}, 3): \citet{hsieh11}, 4): \citet{saka07}, 5): \citet{dav04}, \newline \hspace*{2.5mm} 6): \citet{nak87}, 
7): \citet{weiss99}, 8): \citet{mat00}} \\
\noalign{\smallskip}
 & & [Mpc] & Type      & interaction?         & [pc]        & peak?      & symmetry & dust lane?    & [\#]   &           \\
\noalign{\smallskip}
\hline
\noalign{\smallskip}
NGC~1614 & & 64.2 & SB(s)b pec & + & 231               & - & asym & 1  & 10  & 1) \\
NGC~1097 & & 14.5 & SB(s)b     & - & 700               & + & sym  & 2  & 19  & 2), 3) \\
NGC~1365 & & 17.9 & SB(s)b     & - & 1000              & - & asym & 2  & 5   & 4) \\
NGC~7469 & & 66   & SBa & +\footnote{Part of an interacting galaxy pair with IC~5283.} & 800 & + & asym & -- & -- & 5) \\ 
M~82     & & 3.9  & I0  & +\footnote{Part of an interacting galaxy pair with M~81.}    & 200 & + & sym  & -- & -- & 6)\\
                  & &      &            &   & 105\,$\times$\,70 & - & asym & -- & -- & 7), 8)\\
\noalign{\smallskip}
\hline
\end{tabular}
\end{minipage}
\end{table*}

\subsection{Comparison with other galaxies} \label{subsec:comparison}

A comparison with other galaxies, such as \object{NGC~4194 (the Medusa merger)}, \object{NGC~1097}, and \object{NGC~1365} shows that 
there are similarities between these different types of galaxies and \object{NGC~1614}, but also that NGC~1614 does not fit in either 
in the same class as the \object{Medusa} or the class of barred spiral galaxies.

\subsubsection{Comparison with other dust lane minor mergers} \label{subsubsec:comparison_medusa}

Similar properties as seen in \object{NGC~1614} have been found for the \object{Medusa} merger. The two systems have in common minor axis 
dust lanes and large amounts of molecular gas associated with thesee dust lanes \citep{bes05,ols10,aalto00}, but not with the bulk of the 
star formation.\\
\indent
Apparently, the dust lanes are part of a feeding chain where the majority of the molecular gas has yet not engaged in star formation. 
However, a comparison of the CO distribution and the star formation properties shows a very different picture. in NGC~1614, which is a 
(S+s) minor merger, the compact CO emission is closely associated with the optical nuclear star formation \citep{ols10}, while in 
NGC~4194, an E+S minor merger, much of the star formation is going on in super star clusters (SSCs) with a kpc-scale distribution 
\citep{wei04} away from the molecular gas. No obvious correlation between these young SSCs (5\,$-$\,15~Myr) and the CO can be found \citep{aalto10}. In addition, both minor mergers have a relatively high CO/HCN line ratio 
\citep[CO/HCN\,(NGC~1614)\,$\sim$\,43, CO/HCN\,(\object{NGC~4194})\,$\sim$\,61;][]{costa11}, which indicates limited reservoirs of dense 
gas.\\
\indent
With the Vext data we recover only a fraction of the CO\,(2$-$1) emission in the center of NGC~1614. When we compare our data with the 
lower resolution SMA data by \citet{wil08}, we find that we missed almost all, $\sim$70\%, of the CO\,2$-$1 flux in the crossing minor 
axis dust lane in the north.\\
\indent
Possibly a large fraction of the missing flux is in the form of diffuse extended emission, associated with gas flows along the minor axis 
dust lane. This would be a similar situation as the one suggested for the NGC~4194 minor merger by \citet{aalto10}.\\
\indent
It is still unknown whether the differences in the gas distribution and starbursts are due to different merger histories, variations in 
the process of merging spiral galaxies with spiral galaxies or elliptical with spiral (E+S) galaxies, or diverging properties 
within the evolution of mergers. A careful study of several of these objects might shed some light on their 
evolutionary paths.

\subsubsection{Comparing NGC~1614 with galaxies with nuclear molecular rings} \label{subsubsec:comparison_rings}

\object{NGC~1614} has some features in common with local barred spiral galaxies, such as \object{NGC~1097} and \object{NGC~1365}, but 
unlike NGC~1614, these galaxies are not classified as mergers. These barred spirals are so-called twin-peak galaxies because they contain 
nuclear molecular rings that show a double-peak structure in their molecular gas distribution perpendicular to the stellar bar, where the 
starburst ring and the dust lanes connect \citep{saka07,hsieh11}. NGC~1614 harbors an asymmetric ring and displays at least one 
connection point with its dust lanes. This is comparable to the twin connection points seen in local barred spirals \citep{kenney92}. However, some barred spirals, such as NGC~1365, do show comparable asymmetric central starburst rings \citep{saka07}.\\
\indent
The molecular gas in all three objects is associated with dust lanes close to or connected to the nuclear starburst rings. Within the rings all sources show hot spots in the CO emission that possibly contain SSCs in their early stages \citep{saka07}. An obvious 
difference in the CO gas distribution in the starburst ring though is the absence of CO emission in NGC~1614 at the center of the ring, 
as opposed to the nuclear ring in NGC~1097. This might be an indication that the central structure in NGC~1614 is older than that in 
NGC~1097 and therefore the gas there is already consumed or swept away by developing stars. Another possibility is that the ring and AGN 
activity in NGC~1097 were caused by intermediately dampened spiral density waves \citep{lou03}, whereas the dampening in NGC~1614 was 
strong enough only to form the starburst ring (see also Sect.\,\ref{subsubsec:wild_vs_dyn}).\\
\indent
With sizes (radii) of $\sim$\,700~pc and 1~kpc, the nuclear rings in NGC~1097 and NGC~1365 also have a much larger extension then that of 
NGC~1614 (231~pc), possibly indicating that NGC~1614 harbors a scaled-down version of a nuclear starburst ring in development caused by 
the merger event.\\
\indent
Another prominent galaxy with a molecular ring is \object{M~82}. In contrast to \object{NGC~1614}, M~82 is not a merger, but constitutes 
an integral part of an interacting galaxy pair together with \object{M~81}. M~82 is an edge-on  starburst galaxy with a nuclear molecular 
ring or torus \citep[radius $\sim$200~pc;][]{nak87,shen95} that corresponds to current sites of star formation \citep{nak87}, like in 
NGC~1614. In M~82 in contrast to NGC~1614, however, the star formation is also associated with SSCs \citep{tel92}, like in 
\object{NGC~4194}. The molecular ring in M~82 contains less than 30$\%$ of the total molecular CO mass in M~82. The ring in NGC~1614 
shows a mass fraction of 30$\%$ when taking the total CO\,(2$-$1) mass from \citet{wil08}. In addition, the 
$^{\rm 12}$CO\,/\,$^{\rm 13}$CO\,(2$-$1) ratios for M~82 \citep[$\sim$ 14.2,][]{mao00} and NGC~1614 ($>$15, see 
Sect.\,\ref{subsubsec:13co_12co_ratio}) are quite similar. A more recent study by \citet{weiss99} discovered an additional CO feature in 
M~82 that is suggested to be a superbubble driven by a starburst in its center \citep{mat00}, similar to the wildfire mode proposed by 
\citet{alo01} for NGC~1614.\\
\indent
Yet another example is the Seyfert 1 galaxy \object{NGC~7469}, which is the companion of \object{IC~5283} in an interacting pair of 
similar sized galaxies. For NGC~7469 several tracers (e.g. 5~GHz radio continuum, $J$, $H$, $K$ imaging, PAHs, CO\,(2$-$1)) were found to 
be located in a circumnuclear ring/spiral arm system \citep{wil91,maz94,mil94,gen95,dav04}. A comparison with the starburst and molecular 
rings in NGC~1614 shows some prominent similarities. The near-infrared (NIR) emission detected in the $K$-band continuum, associated with 
the radio continuum and other tracers, is found inside the CO emission distribution \citep{dav04}. For NGC~7469 however, the CO\,(2$-$1) 
distribution shows a central peak/nuclear ring and a bar or pair of spiral arms that seems to connect to the central 
emission/circumnuclear ring itself \citep{dav04}. In NGC~1614 this is not the case.

\subsubsection{GMAs in galaxies with nuclear molecular rings} \label{subsubsec:comparison_rings_gmas}

\noindent
One other property \object{NGC~1614} shares with galaxies like \object{NGC~1097} and \object{NGC~1365} are hot spots 
\citep[gas clumps and CO brightness temperature peaks,][]{saka07,hsieh11}, so-called GMAs \citep{vogel88}. For NGC~1365, \citet{saka07} 
found five CO hot spots with line widths (FWHM) between 60 and 90~km\,s$^{\rm -1}$ that seem to be associated with SSCs of 
$\sim$10$^{\rm 6}$~M$_{\sun}$ \citep{gal05}.\\
\indent
\citet{hsieh11} found many GMAs in the molecular ring (14 GMAs) and the dust lane (5 GMAs) of NGC~1097. They subdivided the GMAs in the 
molecular ring into two categories: GMAs with broad ($\sigma$\,$>$\,30~km\,s$^{\rm -1}$) and narrow ($\sigma$\,$<$\,30~km\,s$^{\rm -1}$) 
line dispersions. The line widths for the narrow GMAs ($\sim$\,55.1~km\,s$^{\rm -1}$) are almost only half the line widths of broad-line 
GMAs ($\sim$\,102.0~km\,s$^{\rm -1}$) but the average molecular gas mass is almost the same 
($M$$_{\rm gas,ave}$(N1-N11)\,$\sim$\,8.1\,$\times$\,10$^{\rm 7}$~M$_{\sun}$, 
$M$$_{\rm gas,ave}$(B1-B3)\,$\sim$\,9.3\,$\times$\,10$^{\rm 7}$~M$_{\sun}$). The GMAs in the dust lane only have slightly more narrow 
line widths ($\sim$\,93.8~km\,s$^{\rm -1}$) and slightly higher molecular gas masses 
($M$$_{\rm gas,ave}$(D1-D5)\,$\sim$\,1.0\,$\times$\,10$^{\rm 8}$~M$_{\sun}$) than the GMAs in the ring. For NGC~1614 the average line 
widths and masses are $\sim$\,129.5~km\,s$^{\rm -1}$ and $M$$_{\rm gas,ave}$(ring)\,$\sim$\,7.0\,$\times$\,10$^{\rm 7}$~M$_{\sun}$ for 
GMAs in the ring, and $\sim$\,110.7~km\,s$^{\rm -1}$ and 
$M$$_{\rm gas,ave}$(outside the ring)\,$\sim$\,5.0\,$\times$\,10$^{\rm 7}$~M$_{\sun}$ for GMAs outside the molecular ring (Table\,\ref{tab:GMA_properties}) - the average line widths are broader and the masses are slightly higher for GMAs in the ring than for those 
outside the ring in NGC~1614. At least two of the GMAs outside the ring in NGC~1614 are associated with dust lanes (GMA\,2, GMA\,5, 
Fig.\,\ref{fig:overlay_co2-1_co1-0_HST}). Three of the GMAs in the ring (GMA\,6, GMA\,7, and GMA\,8) are also associated with a dust 
lane, and GMA\,3 is located in the feeding lines that connect the molecular ring with the dust lane 
(Fig.\,\ref{fig:overlay_co2-1_co1-0_HST}).\\
\indent
\citet{hsieh11} also determined the sizes of their GMAs by measuring the number of pixels above a certain intensity threshold and then 
assuming the GMAs to be spherical to obtain diameters of the individual GMAs. The sizes of GMAs located in the ring of NGC~1097 and 
outside the structure differ significantly. The average size of a GMA situated in the molecular ring is $\sim$220~pc. GMAs that are 
situated outside the ring but in the dust lane in NGC~1097, however, have an average size of 269~pc, which is $\sim$20$\%$ larger than 
for GMAs in the ring. For NGC~1614, however, this seems to be less pronounced. The average size of GMAs in the ring is 
$\sim$259\,$\pm$\,24~pc (assuming spherical morphology) and for GMAs outside the ring and associated with dust lanes it is 
$\sim$273\,$\pm$\,20~pc. The GMAs in the ring in NGC~1614 seem to be larger than those found in NGC~1097 \citep{hsieh11}. Although the 
average spherical sizes for GMAs in and outside the ring are comparable and agree within the error limits, the sizes for the GMAs outside 
the ring that are associated with dust lanes are most likely overestimated. The sizes of these GMAs are measured at low surface 
brightnesses and therefore likely to be afflicted with larger errors. This does not change the fact that \citet{hsieh11} discovered that 
for NGC~1097, GMAs in the ring are significantly smaller than those outside the ring, whereas for NGC~1614 it is most likely the other 
way around, or possibly the sizes in and outside the ring are about the same. One deciding factor here might be the formation history of 
the starburst rings in these two galaxies $-$ e.g., NGC~1614 is a merger, NGC~1097 is not.\\
\indent
We summarize the comparisons between the galaxies in this section in Table\,\ref{tab:vgl_gal_mol_rings}. This shows that NGC~1614 shares 
properties with several of the comparison galaxies - e.g., an assymetric molecular ring, dust lanes, GMAs, and a radial displacement 
between starburst and molecular rings. However, we always found significant differences as well. Therefore, NGC~1614 cannot be classified 
as a twin-peak galaxy, a typical barred spiral, or a typical merger of any flavor; it remains a very interesting yet puzzling object.


\section{Summary} \label{sec:summary}

We have observed the 1.3~mm continuum and the $^{\rm 12}$CO and $^{\rm 13}$CO\,(2$-$1) transitions with high resolution toward 
the center of \object{NGC~1614}.

\begin{itemize}
\item[1.] We detected an asymmetric distribution of the $^{\rm 12}$CO\,(2$-$1) emission in a ring of 231~pc radius with a molecular gas 
mass of 8.3\,$\times$10$^{\rm 8}$~M$_{\sun}$. The center of the ring is free of CO emission.
\item[2.] We did not detect $^{\rm 13}$CO\,(2$-$1) with our observational setup. The lower limit of the 
$^{\rm 12}$CO/$^{\rm 13}$CO\,(2$-$1) line ratio is 15. 
\item[3.] We identified individual GMAs in the CO data cube, some associated with the ring, some associated with the dust lanes. The GMAs 
in the ring have a higher dispersion and more mass than those outside the ring.
\item[4.] Molecular gas in the dust lane shows little evidence of star formation. While GMAs appear in the \textit{umbilical-cord} 
connection between the dust lane and starburst ring, there is no evidence of substantial levels of star formation (e.g., 
Fig.\,\ref{fig:overlay_co2-1_vla_paalpha}).  This behavior is suggestive of a transitory process in which GMAs form as they move toward the nuclear ring, but intense star formation only takes place once the molecular material merges into the ring.
\item[5.] There is no general consensus yet for one scenario to explain the formation of the circumnuclear ring. Considering our results, 
however, we suggest that the starburst ring is caused by a Lindblad resonance and fueled by gas moving via the dust lanes onto the ring.
\end{itemize}
In conclusion, we found that NGC~1614 is a good target for studying the impact of a minor merger on the state of the molecular gas and 
the formation of circumnuclear starburst rings. We proposed a way to explain how nuclear gas structures in minor mergers are fed. 
NGC~1614 may be an example of a merger where gas in tidally induced dust structures has been found to physically connect to a 
starbursting structure in the very nucleus of the galaxy. We also suggested that the GMAs in NGC~1614 are formed in situ in the ring through a combination of cloud-cloud collisions and crowding processes that facilitate the onset of star formation. Owing to the limited 
spatial resolution and surface brightness sensitivity, many questions remain to be answered, such as in which way the larger scale 
CO\,(1$-$0) is connected to the starburst, and if there is a connection to the proposed nuclear outflow \citep{ols10}, or how the 
kinematics of cold molecular gas affect the evolution of star formation and nuclear activity in minor mergers in general. A more detailed 
study of the molecular gas content of NGC~1614 on different spatial scales is needed to address these questions. New studies with 
instruments such as ALMA will enable us to answer them.

\begin{acknowledgements}
  We are grateful to Glen Petitpas for his help with the SMA observations. The Dark Cosmology Centre is funded by the Danish National 
  Research Foundation. SA thanks the Swedish Research Council and the Swedish National Space Board for support. JSG acknowledges partial 
  support for this research from NSF grant AST0708967 from the University of Wisconsin-Madison. The Submillimeter Array is a joint 
  project between the Smithsonian Astrophysical Observatory and the Academia Sinica Institute of Astronomy and Astrophysics and is funded 
  by the Smithsonian Institution and the Academia Sinica. MERLIN is a national facility operated by The University of Manchester on 
  behalf of the Science and Technology Facilities Council (STFC). The VLA is operated by the National Radio Astronomy Observatory, a 
  facility of the National Science foundation operated under cooperative agreement by Associated Universities, Inc. This research has 
  made use of the NASA/IPAC Extragalactic Database (NED) which is operated by the Jet Propulsion Laboratory, California Institute of 
  Technology, under contract with the National Aeronautics and Space Administration. This research used the facilities of the Canadian 
  Astronomy Data Centre operated by the National Research Council of Canada with the support of the Canadian Space Agency.
\end{acknowledgements}

\bibliographystyle{aa}
\bibliography{ngc1614}

\begin{thebibliography}{68}
\expandafter\ifx\csname natexlab\endcsname\relax\def\natexlab#1{#1}\fi

\bibitem[{{Aalto} {et~al.}(2010){Aalto}, {Beswick}, \& {J{\"u}tte}}]{aalto10}
{Aalto}, S., {Beswick}, R., \& {J{\"u}tte}, E. 2010, \aap, 522, A59

\bibitem[{{Aalto} \& {H{\"u}ttemeister}(2000)}]{aalto00}
{Aalto}, S. \& {H{\"u}ttemeister}, S. 2000, \aap, 362, 42

\bibitem[{{Alonso-Herrero} {et~al.}(2001){Alonso-Herrero}, {Engelbracht},
  {Rieke}, {Rieke}, \& {Quillen}}]{alo01}
{Alonso-Herrero}, A., {Engelbracht}, C.~W., {Rieke}, M.~J., {Rieke}, G.~H., \&
  {Quillen}, A.~C. 2001, \apj, 546, 952

\bibitem[{{Alonso-Herrero} {et~al.}(2000){Alonso-Herrero}, {Rieke}, {Rieke}, \&
  {Shields}}]{alo00}
{Alonso-Herrero}, A., {Rieke}, M.~J., {Rieke}, G.~H., \& {Shields}, J.~C. 2000,
  \apj, 530, 688

\bibitem[{{Beswick} {et~al.}(2005){Beswick}, {Aalto}, {Pedlar}, \&
  {H{\"u}ttemeister}}]{bes05}
{Beswick}, R.~J., {Aalto}, S., {Pedlar}, A., \& {H{\"u}ttemeister}, S. 2005,
  \aap, 444, 791

\bibitem[{{Bournaud} {et~al.}(2005){Bournaud}, {Jog}, \& {Combes}}]{bour05}
{Bournaud}, F., {Jog}, C.~J., \& {Combes}, F. 2005, \aap, 437, 69

\bibitem[{{Buta} {et~al.}(1995){Buta}, {van Driel}, {Braine}, {Combes},
  {Wakamatsu}, {Sofue}, \& {Tomita}}]{but95}
{Buta}, R., {van Driel}, W., {Braine}, J., {et~al.} 1995, \apj, 450, 593

\bibitem[{{Chapelon} {et~al.}(1999){Chapelon}, {Contini}, \& {Davoust}}]{cha99}
{Chapelon}, S., {Contini}, T., \& {Davoust}, E. 1999, \aap, 345, 81

\bibitem[{{Combes}(1988{\natexlab{a}})}]{com88b}
{Combes}, F. 1988{\natexlab{a}}, in Lecture Notes in Physics, Berlin Springer
  Verlag, Vol. 315, Molecular Clouds, Milky-Way and External Galaxies, ed.
  R.~L. {Dickman}, R.~L. {Snell}, \& J.~S. {Young}, 441

\bibitem[{{Combes}(1988{\natexlab{b}})}]{com88a}
{Combes}, F. 1988{\natexlab{b}}, in NATO ASIC Proc. 232: Galactic and
  Extragalactic Star Formation, ed. {R.~E.~Pudritz \& M.~Fich}, 475--+

\bibitem[{{Costagliola} {et~al.}(2011){Costagliola}, {Aalto}, {Rodriguez},
  {Muller}, {Spoon}, {Mart{\'{\i}}n}, {Per{\'e}z-Torres}, {Alberdi},
  {Lindberg}, {Batejat}, {J{\"u}tte}, {van der Werf}, \& {Lahuis}}]{costa11}
{Costagliola}, F., {Aalto}, S., {Rodriguez}, M.~I., {et~al.} 2011, \aap, 528,
  A30

\bibitem[{{Dasyra} {et~al.}(2006){Dasyra}, {Tacconi}, {Davies}, {Naab},
  {Genzel}, {Lutz}, {Sturm}, {Baker}, {Veilleux}, {Sanders}, \&
  {Burkert}}]{das06}
{Dasyra}, K.~M., {Tacconi}, L.~J., {Davies}, R.~I., {et~al.} 2006, \apj, 651,
  835

\bibitem[{{Davies} {et~al.}(2004){Davies}, {Tacconi}, \& {Genzel}}]{dav04}
{Davies}, R.~I., {Tacconi}, L.~J., \& {Genzel}, R. 2004, \apj, 602, 148

\bibitem[{{de Vaucouleurs} {et~al.}(1991){de Vaucouleurs}, {de Vaucouleurs},
  {Corwin}, {Buta}, {Paturel}, \& {Fouqu{\'e}}}]{deV91}
{de Vaucouleurs}, G., {de Vaucouleurs}, A., {Corwin}, Jr., H.~G., {et~al.}
  1991, {Third Reference Catalogue of Bright Galaxies. Volume I: Explanations
  and references. Volume II: Data for galaxies between 0$^{h}$ and 12$^{h}$.
  Volume III: Data for galaxies between 12$^{h}$ and 24$^{h}$.}

\bibitem[{{Elmegreen}(1994)}]{elm94}
{Elmegreen}, B.~G. 1994, \apjl, 425, L73

\bibitem[{{Galliano} {et~al.}(2005){Galliano}, {Alloin}, {Pantin}, {Lagage}, \&
  {Marco}}]{gal05}
{Galliano}, E., {Alloin}, D., {Pantin}, E., {Lagage}, P.~O., \& {Marco}, O.
  2005, \aap, 438, 803

\bibitem[{{Genzel} {et~al.}(1995){Genzel}, {Weitzel}, {Tacconi-Garman},
  {Blietz}, {Cameron}, {Krabbe}, {Lutz}, \& {Sternberg}}]{gen95}
{Genzel}, R., {Weitzel}, L., {Tacconi-Garman}, L.~E., {et~al.} 1995, \apj, 444,
  129

\bibitem[{{Haan} {et~al.}(2011){Haan}, {Surace}, {Armus}, {Evans}, {Howell},
  {Mazzarella}, {Kim}, {Vavilkin}, {Inami}, {Sanders}, {Petric}, {Bridge},
  {Melbourne}, {Charmandaris}, {Diaz-Santos}, {Murphy}, {U}, {Stierwalt}, \&
  {Marshall}}]{haan11}
{Haan}, S., {Surace}, J.~A., {Armus}, L., {et~al.} 2011, \aj, 141, 100

\bibitem[{{Hirota} {et~al.}(2011){Hirota}, {Kuno}, {Sato}, {Nakanishi},
  {Tosaki}, \& {Sorai}}]{hiro11}
{Hirota}, A., {Kuno}, N., {Sato}, N., {et~al.} 2011, \apj, 737, 40

\bibitem[{{Ho}(1999)}]{ho99}
{Ho}, L.~C. 1999, Advances in Space Research, 23, 813

\bibitem[{{Hsieh} {et~al.}(2008){Hsieh}, {Matsushita}, {Lim}, {Kohno}, \&
  {Sawada-Satoh}}]{hsieh08}
{Hsieh}, P.-Y., {Matsushita}, S., {Lim}, J., {Kohno}, K., \& {Sawada-Satoh}, S.
  2008, \apj, 683, 70

\bibitem[{{Hsieh} {et~al.}(2011){Hsieh}, {Matsushita}, {Liu}, {Ho}, {Oi}, \&
  {Wu}}]{hsieh11}
{Hsieh}, P.-Y., {Matsushita}, S., {Liu}, G., {et~al.} 2011, \apj, 736, 129

\bibitem[{{Hunsberger} {et~al.}(1996){Hunsberger}, {Charlton}, \&
  {Zaritsky}}]{hun96}
{Hunsberger}, S.~D., {Charlton}, J.~C., \& {Zaritsky}, D. 1996, \apj, 462, 50

\bibitem[{{Jesseit} {et~al.}(2007){Jesseit}, {Naab}, {Peletier}, \&
  {Burkert}}]{jes07}
{Jesseit}, R., {Naab}, T., {Peletier}, R.~F., \& {Burkert}, A. 2007, \mnras,
  376, 997

\bibitem[{{Kenney} {et~al.}(1992){Kenney}, {Wilson}, {Scoville}, {Devereux}, \&
  {Young}}]{kenney92}
{Kenney}, J.~D.~P., {Wilson}, C.~D., {Scoville}, N.~Z., {Devereux}, N.~A., \&
  {Young}, J.~S. 1992, \apjl, 395, L79

\bibitem[{{Knapen} {et~al.}(2004){Knapen}, {Whyte}, {de Blok}, \& {van der
  Hulst}}]{kna04}
{Knapen}, J.~H., {Whyte}, L.~F., {de Blok}, W.~J.~G., \& {van der Hulst}, J.~M.
  2004, \aap, 423, 481

\bibitem[{{Koda} {et~al.}(2009){Koda}, {Scoville}, {Sawada}, {La Vigne},
  {Vogel}, {Potts}, {Carpenter}, {Corder}, {Wright}, {White}, {Zauderer},
  {Patience}, {Sargent}, {Bock}, {Hawkins}, {Hodges}, {Kemball}, {Lamb},
  {Plambeck}, {Pound}, {Scott}, {Teuben}, \& {Woody}}]{koda09}
{Koda}, J., {Scoville}, N., {Sawada}, T., {et~al.} 2009, \apjl, 700, L132

\bibitem[{{Kormendy} {et~al.}(2009){Kormendy}, {Fisher}, {Cornell}, \&
  {Bender}}]{kor09}
{Kormendy}, J., {Fisher}, D.~B., {Cornell}, M.~E., \& {Bender}, R. 2009, \apjs,
  182, 216

\bibitem[{{Lou}(2003)}]{lou03}
{Lou}, Y.~Q. 2003, Acta Astronomica Sinica, 44, 172

\bibitem[{{Lou} {et~al.}(2001{\natexlab{a}}){Lou}, {Yuan}, \& {Fan}}]{lou01a}
{Lou}, Y.-Q., {Yuan}, C., \& {Fan}, Z. 2001{\natexlab{a}}, \apj, 552, 189

\bibitem[{{Lou} {et~al.}(2001{\natexlab{b}}){Lou}, {Yuan}, {Fan}, \&
  {Leon}}]{lou01b}
{Lou}, Y.-Q., {Yuan}, C., {Fan}, Z., \& {Leon}, S. 2001{\natexlab{b}}, \apjl,
  553, L35

\bibitem[{{Mao} {et~al.}(2000){Mao}, {Henkel}, {Schulz}, {Zielinsky},
  {Mauersberger}, {St{\"o}rzer}, {Wilson}, \& {Gensheimer}}]{mao00}
{Mao}, R.~Q., {Henkel}, C., {Schulz}, A., {et~al.} 2000, \aap, 358, 433

\bibitem[{{Matsushita} {et~al.}(2000){Matsushita}, {Kawabe}, {Matsumoto},
  {Tsuru}, {Kohno}, {Morita}, {Okumura}, \& {Vila-Vilar{\'o}}}]{mat00}
{Matsushita}, S., {Kawabe}, R., {Matsumoto}, H., {et~al.} 2000, \apjl, 545,
  L107

\bibitem[{{Mazzarella} {et~al.}(1994){Mazzarella}, {Voit}, {Soifer},
  {Matthews}, {Graham}, {Armus}, \& {Shupe}}]{maz94}
{Mazzarella}, J.~M., {Voit}, G.~M., {Soifer}, B.~T., {et~al.} 1994, \aj, 107,
  1274

\bibitem[{{Mazzuca} {et~al.}(2006){Mazzuca}, {Sarzi}, {Knapen}, {Veilleux}, \&
  {Swaters}}]{maz06}
{Mazzuca}, L.~M., {Sarzi}, M., {Knapen}, J.~H., {Veilleux}, S., \& {Swaters},
  R. 2006, \apjl, 649, L79

\bibitem[{{Meier} \& {Turner}(2004)}]{mei04}
{Meier}, D.~S. \& {Turner}, J.~L. 2004, \aj, 127, 2069

\bibitem[{{Miles} {et~al.}(1994){Miles}, {Houck}, \& {Hayward}}]{mil94}
{Miles}, J.~W., {Houck}, J.~R., \& {Hayward}, T.~L. 1994, \apjl, 425, L37

\bibitem[{{Monreal-Ibero} {et~al.}(2006){Monreal-Ibero}, {Arribas}, \&
  {Colina}}]{mon06}
{Monreal-Ibero}, A., {Arribas}, S., \& {Colina}, L. 2006, \apj, 637, 138

\bibitem[{{Nakai} {et~al.}(1987){Nakai}, {Hayashi}, {Handa}, {Sofue},
  {Hasegawa}, \& {Sasaki}}]{nak87}
{Nakai}, N., {Hayashi}, M., {Handa}, T., {et~al.} 1987, \pasj, 39, 685

\bibitem[{{Narayanan} {et~al.}(2011){Narayanan}, {Krumholz}, {Ostriker}, \&
  {Hernquist}}]{nar11}
{Narayanan}, D., {Krumholz}, M., {Ostriker}, E.~C., \& {Hernquist}, L. 2011,
  \mnras, 418, 664

\bibitem[{{Neff} {et~al.}(1990){Neff}, {Hutchings}, {Standord}, \&
  {Unger}}]{neff90}
{Neff}, S.~G., {Hutchings}, J.~B., {Standord}, S.~A., \& {Unger}, S.~W. 1990,
  \aj, 99, 1088

\bibitem[{{Olsson} {et~al.}(2010){Olsson}, {Aalto}, {Thomasson}, \&
  {Beswick}}]{ols10}
{Olsson}, E., {Aalto}, S., {Thomasson}, M., \& {Beswick}, R. 2010, \aap, 513,
  A11+

\bibitem[{{Puxley} \& {Brand}(1999)}]{pux99}
{Puxley}, P.~J. \& {Brand}, P.~W.~J.~L. 1999, \apj, 514, 675

\bibitem[{{Sakamoto} {et~al.}(2007){Sakamoto}, {Ho}, {Mao}, {Matsushita}, \&
  {Peck}}]{saka07}
{Sakamoto}, K., {Ho}, P.~T.~P., {Mao}, R.-Q., {Matsushita}, S., \& {Peck},
  A.~B. 2007, \apj, 654, 782

\bibitem[{{Sanders} {et~al.}(2003){Sanders}, {Mazzarella}, {Kim}, {Surace}, \&
  {Soifer}}]{san03}
{Sanders}, D.~B., {Mazzarella}, J.~M., {Kim}, D.-C., {Surace}, J.~A., \&
  {Soifer}, B.~T. 2003, \aj, 126, 1607

\bibitem[{{Sarzi} {et~al.}(2007){Sarzi}, {Allard}, {Knapen}, \&
  {Mazzuca}}]{sar07}
{Sarzi}, M., {Allard}, E.~L., {Knapen}, J.~H., \& {Mazzuca}, L.~M. 2007,
  \mnras, 380, 949

\bibitem[{{Scoville} {et~al.}(1986){Scoville}, {Sanders}, \& {Clemens}}]{sco86}
{Scoville}, N.~Z., {Sanders}, D.~B., \& {Clemens}, D.~P. 1986, \apjl, 310, L77

\bibitem[{{Scoville} {et~al.}(1985){Scoville}, {Soifer}, {Neugebauer},
  {Matthews}, {Young}, \& {Yerka}}]{sco85}
{Scoville}, N.~Z., {Soifer}, B.~T., {Neugebauer}, G., {et~al.} 1985, \apj, 289,
  129

\bibitem[{{Shen} \& {Lo}(1995)}]{shen95}
{Shen}, J. \& {Lo}, K.~Y. 1995, \apjl, 445, L99

\bibitem[{{Shlosman} {et~al.}(1989){Shlosman}, {Frank}, \& {Begelman}}]{shlo89}
{Shlosman}, I., {Frank}, J., \& {Begelman}, M.~C. 1989, \nat, 338, 45

\bibitem[{{Simkin} {et~al.}(1980){Simkin}, {Su}, \& {Schwarz}}]{sim80}
{Simkin}, S.~M., {Su}, H.~J., \& {Schwarz}, M.~P. 1980, \apj, 237, 404

\bibitem[{{Somerville} {et~al.}(2008){Somerville}, {Hopkins}, {Cox},
  {Robertson}, \& {Hernquist}}]{som08}
{Somerville}, R.~S., {Hopkins}, P.~F., {Cox}, T.~J., {Robertson}, B.~E., \&
  {Hernquist}, L. 2008, \mnras, 391, 481

\bibitem[{{Stutzki} \& {Guesten}(1990)}]{stu90}
{Stutzki}, J. \& {Guesten}, R. 1990, \apj, 356, 513

\bibitem[{{Tacconi} {et~al.}(2002){Tacconi}, {Genzel}, {Lutz}, {Rigopoulou},
  {Baker}, {Iserlohe}, \& {Tecza}}]{tac02}
{Tacconi}, L.~J., {Genzel}, R., {Lutz}, D., {et~al.} 2002, \apj, 580, 73

\bibitem[{{Tan}(2000)}]{tan00}
{Tan}, J.~C. 2000, \apj, 536, 173

\bibitem[{{Telesco} \& {Gezari}(1992)}]{tel92}
{Telesco}, C.~M. \& {Gezari}, D.~Y. 1992, \apj, 395, 461

\bibitem[{{Terashima} {et~al.}(2000){Terashima}, {Ho}, {Ptak}, {Mushotzky},
  {Serlemitsos}, {Yaqoob}, \& {Kunieda}}]{tera00}
{Terashima}, Y., {Ho}, L.~C., {Ptak}, A.~F., {et~al.} 2000, \apj, 533, 729

\bibitem[{{Toomre} \& {Toomre}(1972)}]{too72}
{Toomre}, A. \& {Toomre}, J. 1972, \apj, 178, 623

\bibitem[{{Toth} \& {Ostriker}(1992)}]{toth92}
{Toth}, G. \& {Ostriker}, J.~P. 1992, \apj, 389, 5

\bibitem[{{V{\"a}is{\"a}nen} {et~al.}(2012){V{\"a}is{\"a}nen}, {Rajpaul},
  {Zijlstra}, {Reunanen}, \& {Kotilainen}}]{vai12}
{V{\"a}is{\"a}nen}, P., {Rajpaul}, V., {Zijlstra}, A.~A., {Reunanen}, J., \&
  {Kotilainen}, J. 2012, \mnras, 420, 2209

\bibitem[{{Veilleux} {et~al.}(1999){Veilleux}, {Sanders}, \& {Kim}}]{vei99}
{Veilleux}, S., {Sanders}, D.~B., \& {Kim}, D.-C. 1999, \apj, 522, 139

\bibitem[{{Verdes-Montenegro} {et~al.}(1995){Verdes-Montenegro}, {Bosma}, \&
  {Athanassoula}}]{ver95}
{Verdes-Montenegro}, L., {Bosma}, A., \& {Athanassoula}, E. 1995, \aap, 300, 65

\bibitem[{{Vogel} {et~al.}(1988){Vogel}, {Kulkarni}, \& {Scoville}}]{vogel88}
{Vogel}, S.~N., {Kulkarni}, S.~R., \& {Scoville}, N.~Z. 1988, \nat, 334, 402

\bibitem[{{Wei{\ss}} {et~al.}(1999){Wei{\ss}}, {Walter}, {Neininger}, \&
  {Klein}}]{weiss99}
{Wei{\ss}}, A., {Walter}, F., {Neininger}, N., \& {Klein}, U. 1999, \aap, 345,
  L23

\bibitem[{{Weistrop} {et~al.}(2004){Weistrop}, {Eggers}, {Hancock}, {Nelson},
  {Bachilla}, \& {Kaiser}}]{wei04}
{Weistrop}, D., {Eggers}, D., {Hancock}, M., {et~al.} 2004, \aj, 127, 1360

\bibitem[{{Wilson} {et~al.}(1991){Wilson}, {Helfer}, {Haniff}, \&
  {Ward}}]{wil91}
{Wilson}, A.~S., {Helfer}, T.~T., {Haniff}, C.~A., \& {Ward}, M.~J. 1991, \apj,
  381, 79

\bibitem[{{Wilson} {et~al.}(2008){Wilson}, {Petitpas}, {Iono}, {Baker}, {Peck},
  {Krips}, {Warren}, {Golding}, {Atkinson}, {Armus}, {Cox}, {Ho}, {Juvela},
  {Matsushita}, {Mihos}, {Pihlstrom}, \& {Yun}}]{wil08}
{Wilson}, C.~D., {Petitpas}, G.~R., {Iono}, D., {et~al.} 2008, \apjs, 178, 189

\bibitem[{{Zaritsky}(1995)}]{zar95}
{Zaritsky}, D. 1995, \apjl, 448, L17

\end{thebibliography}

\end{document}